# Composition dependence of the critical Rayleigh number curve for macrosegregation in multicomponent metal alloys

O.S. Houghton[1,*], A.S. Sabau[2], G.B. Olson[1,3]

[1]Materials Research Laboratory, Massachusetts Institute of Technology, 77 Massachusetts Avenue, Cambridge, MA, 02139, USA

[2]Computational Sciences and Engineering Division, Oak Ridge National Laboratory, Oak Ridge, Tennessee, 37831, USA

[3]Dept. Materials Science & Engineering, Massachusetts Institute of Technology, 77 Massachusetts Avenue, Cambridge, MA, 02139, USA

*Corresponding author e-mail address: oshough@mit.edu

## Abstract

Convective instabilities in the semi-solid mushy zone can trigger channel formation that leads to defects known as freckles, channel segregates and A-type segregates. In the present work, Flemings' model is used to determine conditions for the onset of local remelting when buoyancy-driven flow is suddenly triggered in an initially stagnant mushy zone. An expression in the form of a Rayleigh number $Ra$, and its associated critical value $Ra_{\text{crit}}$, above which the local remelting condition is satisfied, are derived. Using thermophysical data from CALPHAD, these expressions are evaluated using results from benchmark experimental and numerical studies for the nickel-based superalloy SX-1 and Pb–Sn alloys. The correlation of this local-remelting criterion with previously reported empirical criteria is also tested for various steel compositions. It is found that $Ra_{\text{crit}}$ varies with the local average solid fraction and several thermophysical properties. Since these properties can vary substantially within a relatively narrow composition range, it is suggested that $Ra_{\text{crit}}$ is a strongly composition-dependent parameter.

## 1. Introduction

Casting remains an important processing technique for producing large ingots and small components requiring specific microstructures, such as single-crystal turbine blades [1,2]. A common issue in casting is macrosegregation, first reported by Biringuccio in 1540 [3], which refers to chemical and structural variations on the length scale of a casting. Defects of this kind create challenges in subsequent processing steps, and can lead to premature failure of components during their use. Consequently, castings displaying macrosegregation are typically rejected [1,2].

A range of different macrosegregation defects have been reported [4], of which A-type segregate formation in large steel ingots and freckle formation in directionally solidified Ni-based superalloys are particularly prominent and problematic [1,2,5]. Despite their different names, these defects form by the same mechanism. They are caused by outward buoyancy-driven flow of solute-enriched liquid from the solid-mush interface [6–10]. When the interdendritic flow is sufficiently strong, cold solute-enriched liquid begins to arrest solidification and even remelt outer regions of the mushy zone. The resulting channels, also known as *chimneys*, allow plumes of solute-enriched liquid to flow into the bulk liquid and lead to compositional variation across the entire casting. The channels subsequently freeze with near-eutectic composition [6–9], and are identifiable in the as-cast microstructure as "freckles" in Ni-based superalloys, when they contain spurious grains [11,12], and "channel" or "A-type" segregates in steel ingots.

To describe the conditions associated with channel formation during casting, empirical criteria in terms of the local thermal gradient $G$ and velocity of the solidification isotherms $R$ have been proposed. These criteria, based on functions of $G$ and $R$ for which channel formation occurs below a critical value, have been developed for particular alloy systems [5, 13–16]. Although they successfully outline the general dependencies of channel formation on casting conditions, they have limited use for understanding composition-dependent effects or the design of casting alloys; they apply only to the studied casting geometries and compositions, and combine multiple effects into a single empirical parameter, whose uncertainty is determined by practical limits on experimental precision.

Criteria developed from models that quantitatively describe the mechanism leading to the onset of channel formation offer a more promising route for the design macrosegregation-resistant alloys and defect-free castings. Such criteria can help to reveal the dependence on thermophysical properties, which makes them effective parametric constraints in the design of alloys and components within a concurrent engineering workflow [17,18].

Flemings *et al.* first proposed a heuristic model to describe the instability of the mushy zone due to local solute redistribution [6,10,19,20]. In the original model, local remelting that may lead to channel formation is found to occur when the outward flow of liquid from the mushy zone exceeds the velocity of the advancing solidification front. This criterion has been successfully and widely applied to numerical simulations of casting, but recent challenges of this criterion's ability to generally describe the onset of freckling motivates its re-evaluation [21,22].

A second approach focuses on describing the stability of an initially stagnant mushy zone using a modified Rayleigh number ($Ra$). Originally derived using a formal stability argument [8,9,23–25], Worster studied the scenario of vertical directional solidification with a planar solidification front under near-eutectic and large far-field temperature asymptotic limits [26–28]. $Ra$ describes the force imbalance between buoyancy and resistance to flow by the porous mushy zone. When the force imbalance is sufficiently large, these formal analyses show that various complex modes of convection can theoretically occur due to buoyancy-driven growth of perturbations to the solidification front in a non-convecting mushy-layer base state (stagnant mushy zone). The onset of natural convection may occur as a result of infinitesimal perturbations of the mushy zone (supercritical bifurcation) when $Ra$ is sufficiently large, or from perturbations of finite amplitude (subcritical bifurcation) at lower values of $Ra$ [8,9]. Under the assumed asymptotic limits, expressions for the critical values of $Ra$ above which the onset of natural convection occurs in response to each kind of perturbation have been determined [26,27].

A definition of $Ra$ has been adopted in physical metallurgy, but the critical values for each kind of instability have not been derived [1,2,29]. Experimental studies have instead sought to empirically determine a value for $Ra_{\mathrm{crit}}$, by computing the value of $Ra$ at which the channels are observed. These studies suggest that, in the absence of variations related to specific casting conditions, there is a single $Ra_{\mathrm{crit}}$ value that is constant for a particular alloy family. Accordingly, $Ra_{\mathrm{crit}}$ values have been reported for each alloy system [1,2], but these values are relatively uncertain and so a $Ra$-based approach currently has limited effectiveness for alloy design or selection.

Efforts to more precisely determine $Ra_{\mathrm{crit}}$ have centered on using numerical process simulations—continuum model simulations of an entire casting that solve the coupled differential equations which describe the temperature, velocity, pressure and composition fields [30]. Such studies have suggested modifications and scaling factors for $Ra$ to account for particle movement, the geometry of the casting, and the orientation of the solidification front with respect to gravity [1,19,31]. Unfortunately, both process and microscale simulations of the required precision to accurately predict the onset of channel formation are computationally expensive [4,10,22,32]. These demands on computational power consequently constrain each simulation study to a small number of casting geometries and compositions, and suggest an alloy discovery approach based on a high-throughput screening approach of casting simulations is impractical. The dependence of $Ra$ and $Ra_{\mathrm{crit}}$ on thermophysical properties of the alloy, and thus the ability to prevent channel formation through optimization of alloy composition, therefore remains unclear.

In the present work, Flemings' local remelting criterion is expressed in a Rayleigh-number form. Using an initially stagnant mushy zone as the reference state, analogous to the non-convecting base state considered by Worster, we obtain expressions for the critical value above which the local remelting condition is satisfied. We test this model by computing $Ra$ and $Ra_{\mathrm{crit}}$ curves for a range of compositions across different alloy families whose solidification behavior has been studied experimentally and numerically. Our re-expression of Flemings' criterion allows us to describe an underlying mechanism that can lead to channel formation; compare Flemings' criterion with $Ra$-based approaches; describe the effect of composition on susceptibility to local remelting; and present $Ra$ and $Ra_{\mathrm{crit}}$ in a form which can be computed using CALPHAD, and therefore appears practical to assist alloy design. The limitations of this model, with respect to its simplifying assumptions, non-linear effects, and the role of defects, are also briefly discussed.

## 2. Methods

For Ni-based superalloys, thermophysical properties and composition and temperature profiles during solidification were calculated using the TCNI Nickel-based Superalloys Database version 12 (TCNI12) [33] and the Scheil-Gulliver model in Thermo-Calc (Version 2024b) [34–36]. For steels, equivalent calculations were performed using QuesTek's ICMD toolkit, Computherm's Pandat software, and the PanIron 2025 database [37–39]. Carbon was assumed to act as a “fast diffuser” (complete diffusion in the solid), and each composition was constrained to solidify into only cubic close-packed Austenite and body-centred cubic Ferrite.

## **3. Theory**

### *3.1 Re-expression of Flemings’ criterion for mushy zone stability*

The group of Flemings outlined a model for the onset of channel formation during solidification of a multicomponent alloy [19,20]. They considered a small volume element within an isotropic and uniformly varying mushy zone. This volume element is large enough for the fraction solid within it at any time to represent the local average but small enough that it can be treated as a differential element during solidification at the advancing planar front (Figure 1).

It is also assumed that the solid and liquid phases are in local thermodynamic equilibrium at the interface between them, such that the liquid composition is related to the liquidus temperature; no solid material enters or leaves the volume element during solidification; solute enters or leaves the element only by liquid flow to feed shrinkage solidification and expansion or contraction of the liquid with temperature and composition; mass within the differential element is conserved; the density of the solid is constant; there is no pore formation; and the liquid composition and temperature within the volume element are uniform within a differential amount at time *t*.

For solutes that are substitutional, there is perfect mixing in the liquid but no diffusion in the solid. The composition of the liquid then varies with time according to the “local solute redistribution equation” [19]:

$$\frac{\partial C_{\mathrm{L}}^{i}}{\partial t} = -\left(\frac{1-k^{i}}{1-\beta}\right)\frac{C_{\mathrm{L}}^{i}}{f_{\mathrm{L}}}\frac{\partial f_{\mathrm{L}}}{\partial t} - \boldsymbol{v}\cdot\boldsymbol{\nabla}C_{\mathrm{L}}^{i}. \tag{1}$$

For solutes that are interstitial, such as carbon in steels, there is perfect mixing in the liquid and complete diffusion in the solid [20]:

$$\frac{\partial C_{\mathrm{L}}^{j}}{\partial t} = -\frac{C_{\mathrm{L}}^{j}(1-k^{j})}{k^{j}(1-f_{\mathrm{L}})+f_{L}(1-\beta)}\frac{\partial f_{L}}{\partial t} - \boldsymbol{v}\cdot\boldsymbol{\nabla}C_{\mathrm{L}}^{j}. \tag{2}$$

$C_{\mathrm{L}}$ is the composition of the liquid, $k^{i}$ is the partition coefficient of solute element $i$, $\beta$ is the solidification shrinkage given by $(\rho_{S}-\rho_{L})/\rho_{S}$ where $\rho_{\mathrm{S}}$ is the density of the solid and $\rho_{\mathrm{L}}$ is the density of the liquid. $f_L$ is the volumetric liquid fraction, and $\boldsymbol{v}$ is the inter-dendritic velocity of the liquid relative to the solid.

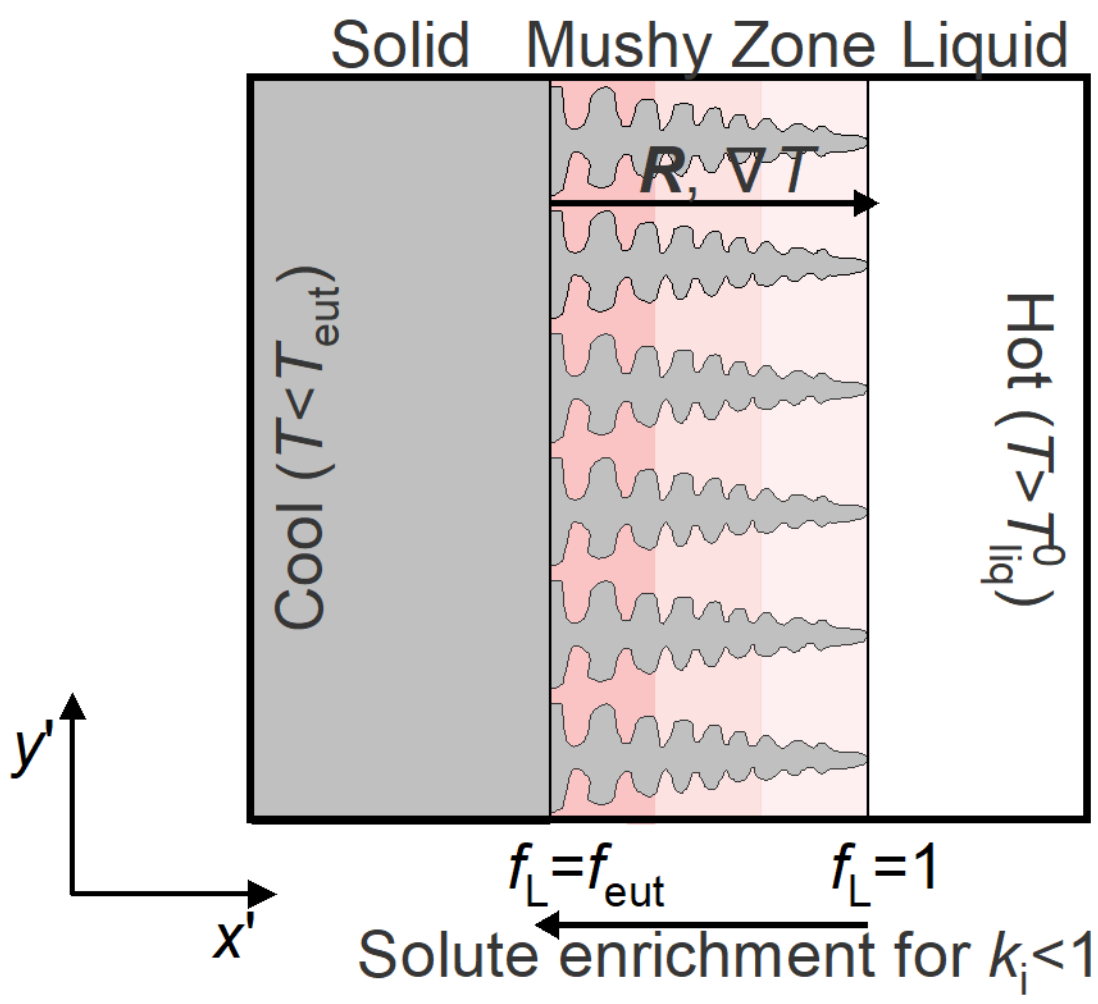

**Figure 1**: Schematic of solidification in the mushy zone, where $f_L$ represents the fraction of liquid, $f_{eut}$ indicates the fraction of residual liquid when it reaches the eutectic composition or otherwise fully solidifies. Red shading indicates the degree of solute enrichment in the liquid for element $i$ with a partition coefficient $k_i$ less than one. $T_{eut}$ is the eutectic temperature, and $T^0_{liq}$ is the liquidus temperature of the nominal composition. $\boldsymbol{R}$ is the velocity of the solidification isotherms, and $\boldsymbol{\nabla} T$ is the temperature gradient. Dendrites are shown for illustrative purposes only.

Since we assume that the liquid is in equilibrium with the solid at their interface, the temperature and composition field of the liquid are coupled. The temperature of the volume element is then the liquidus temperature $T_{liq}$ corresponding to the solid fraction present. This can be related to the composition of the liquid:

$$T_{\mathrm{liq}} = T'_{\mathrm{liq}} + \sum_n m_{\mathrm{L},n} C_{\mathrm{L},n}, \tag{3}$$

where $T'_{liq}$ is a constant and $m_{L,n}$ is the liquidus slope with respect to solute element $n$. Eq. 3 assumes a linear liquidus relationship which is reasonable for small changes in composition. Combining Eqs. 1–3 yields:

$$\frac{\partial f_L}{\partial T} = \frac{1+(\boldsymbol{v}\cdot\boldsymbol{\nabla}T)\left(\frac{\partial T}{\partial t}\right)^{-1}}{-\sum_i m_{\mathrm{L},i}\left(\frac{1-k^i}{1-\beta}\right)\frac{C_{\mathrm{L}}^i}{f_{\mathrm{L}}}-\sum_j m_{\mathrm{L},j}\frac{C_{\mathrm{L}}^j\left(1-k^j\right)}{k^j(1-f_{\mathrm{L}})+f_{\mathrm{L}}(1-\beta)}}. \tag{4}$$

Eq. 4 describes the change in liquid fraction with respect to temperature, but the onset of freckling is dependent on both position and time. The change in the liquid fraction with respect to time requires consideration of heat transfer across the volume element [22]:

$$(1-\beta f_{\mathrm{L}})\frac{\partial T}{\partial t}=\alpha\nabla^{2}T-\frac{H}{C_{\mathrm{p}}}\frac{\partial f_{\mathrm{L}}}{\partial t}-(1-\beta)f_{\mathrm{L}}\boldsymbol{v}\cdot\boldsymbol{\nabla}T. \tag{5}$$

$H$ is the latent heat of fusion, α is the thermal diffusivity, and $C_{\mathrm{p}}$ is the specific heat capacity of the volume element. Eq. 5 accounts for advection, conduction, and latent heat due to solidification. Combining Eqs. 4 and 5 gives:

$$\frac{\partial f_{\mathrm{L}}}{\partial t}=\frac{\alpha\nabla^{2}T+(1-f_{\mathrm{L}})(\boldsymbol{v}\cdot\boldsymbol{\nabla}T)}{(1-\beta f_{\mathrm{L}})\left[-\sum_{i}m_{\mathrm{L},i}\left(\frac{1-k_{i}}{1-\beta}\right)\frac{C_{\mathrm{L}}^{i}}{f_{\mathrm{L}}}-\sum_{j}m_{\mathrm{L},j}\frac{\left(1-k_{j}\right)C_{\mathrm{L}}^{j}}{k_{j}(1-f_{\mathrm{L}})+f_{L}(1-\beta)}\right]+\frac{H}{C_{\mathrm{p}}}}. \tag{6}$$

When $\partial f_{\mathrm{L}}/\partial t$ is positive, the local remelting condition, which is associated with the initiation of a channel in the mushy zone [20], is satisfied. Since the denominator on the right-hand side of Eq. 6 is always positive, the local remelting condition is satisfied when:

$$\alpha\nabla^{2}T+(1-f_{\mathrm{L}})(\boldsymbol{v}\cdot\boldsymbol{\nabla}T)>0. \tag{7}$$

This result was previously reported in ref. [20]. Inserting Eq. 5 into Eq. 7 gives:

$$(\boldsymbol{v}\cdot\boldsymbol{\nabla}T)>-\dot{T}\left[1+\frac{1}{(1-\beta f_{\mathrm{L}})}\frac{H}{C_{\mathrm{p}}\dot{T}}\frac{\partial f_{\mathrm{L}}}{\partial t}\right]. \tag{8}$$

The present work considers local remelting ($\partial f_{\mathrm{L}}/\partial t>0$) on cooling ($\dot{T}<0$), such that the condition is satisfied when ($\partial f_{\mathrm{L}}/\partial T<0$). Eq. 8 can therefore be expressed as:

$$\frac{(\boldsymbol{v}\cdot\boldsymbol{\nabla}T)}{\dot{T}}<-W\left(\frac{H}{C_{\mathrm{p}}},\frac{1}{\dot{T}}\frac{\partial f_{\mathrm{L}}}{\partial t},f_{\mathrm{L}},\beta\right), \tag{9}$$

where the function $W$ is given by:

$$W\left(H,C_{\mathrm{p}},\frac{1}{\dot{T}}\frac{\partial f_{\mathrm{L}}}{\partial t},f_{\mathrm{L}},\beta\right)=\left[1+\frac{H}{C_{\mathrm{p}}}\frac{1}{(1-\beta f_{\mathrm{L}})}\frac{1}{\dot{T}}\frac{\partial f_{\mathrm{L}}}{\partial t}\right]. \tag{10}$$

Eqs. 9 and 10 are an expression of Flemings' original criterion that explicitly describe the role of heat required to remelt the solidifying material in the mushy zone. At the limiting condition for arrest of local solidification, $\boldsymbol{v}\cdot\boldsymbol{\nabla}T=-\dot{T}$ such that $\partial f_{\mathrm{L}}/\partial t=0$, the function $W$ is equal to 1, and Eq. 9 reduces to the classical criterion. The function $W$ therefore does not modify the condition for local remelting at its onset; it retains the latent heat contribution so it can be re-expressed in a Rayleigh-number form.

*3.2 Derivation of the critical Rayleigh number for upward directional solidification*

Worster derived $Ra$ for upward directional solidification of an alloy [8,9,23–25]. In this scenario, the onset of natural convection can occur when a liquid cooled from below solidifies to leave a less dense residual liquid (Figure 2). In Worster's analysis, $Ra^{W}_{\mathrm{crit}}$ describes the value of $Ra$ above which the onset of natural convection occurs due to a loss of linear stability in a non-convecting mushy-layer base state (stagnant mushy zone).

To consider a scenario analogous to the one in Worster's model, we consider a mushy zone in which the liquid is initially stagnant ($\boldsymbol{v} \cdot \boldsymbol{\nabla} T = 0$). Generally, the value of $\boldsymbol{v} \cdot \boldsymbol{\nabla} T$ and the solidification rate $-\partial f_{\mathrm{L}}/\partial t$ are coupled, so their exact values must be determined numerically. However, in this stagnant scenario, the composition of the liquid is related only to the temperature of the local-average volume element. As a result, $\dot{T}^{-1}\left(\frac{\partial f_L}{\partial t}\right)$ is to good approximation given by $\partial f_{\mathrm{L}}/\partial T_{\mathrm{liq}}$ for the unperturbed solidification path. $\partial f_{\mathrm{L}}/\partial T_{\mathrm{liq}}$ represents the change in liquid fraction with liquidus temperature, which can be determined by considering phase equilibria.

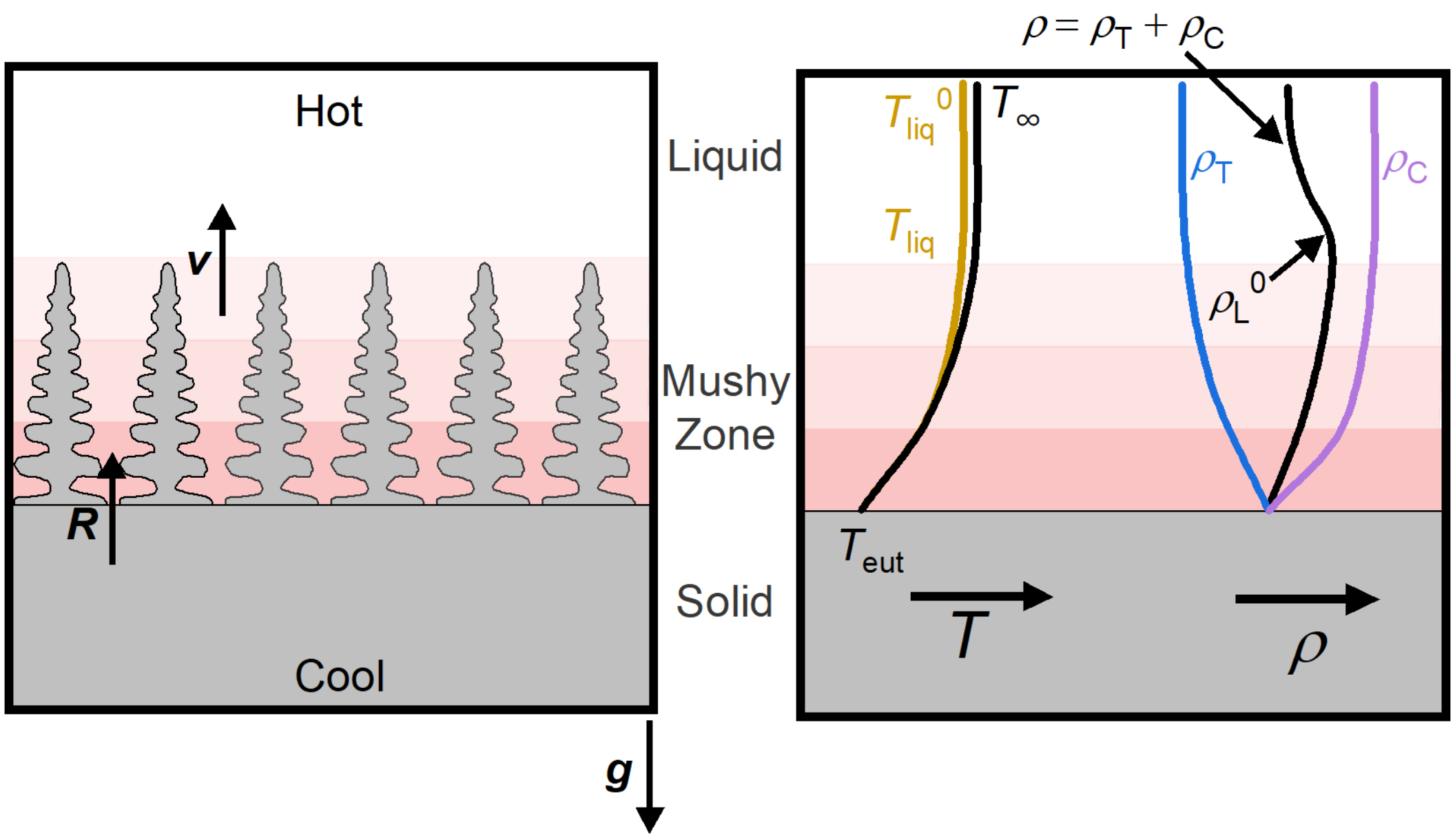

**Figure 2.** Schematics of solidification when a liquid is cooled from below. A horizontal solidification front advances upwards, at speed $R$, to leave a less dense liquid. The solid and liquid are in thermodynamic equilibrium at the interface between them such that the temperature and composition of the liquid are coupled. Moving upwards, away from the solid-mush interface, the liquid becomes hotter and less solute-rich. The opposing effects of temperature and composition on density (right) are described by $\rho_{\mathrm{T}}$ and $\rho_{\mathrm{C}}$, respectively. The liquid ultimately becomes more dense through the mushy zone. When the composition in the fully liquid region reaches the nominal composition, the liquid has density $\rho_L^0$ and is at temperature $T_{liq}^0$. A small compositional boundary layer exists, but its effect is suppressed by the assumptions made in this model. The schematic temperature and density profiles, shown right, are redrawn from Worster [9].

We then assume that the system solidifies into a single solid phase, which is reasonable to assume for the early stages of solidification in which channel formation occurs [2,9,20]. For a Boussinesq liquid, the velocity can be appropriately described using the equation for Darcy flow through a porous medium:

$$\boldsymbol{v} = -\frac{\bar{K}}{\mu f_{\mathrm{L}}}(\boldsymbol{\nabla} P + \rho_{\mathrm{L}}\boldsymbol{g}), \tag{11}$$

where $\bar{K}$ is the mean permeability of the mushy zone in the volume element, $\mu$ is the dynamic viscosity of the fluid, $P$ is the pressure of the fluid and $\boldsymbol{g}$ has its usual meaning [20]. From Eq. 9, when a buoyancy-driven flow is suddenly triggered, local remelting of the reference stagnant mushy zone occurs when:

$$\Delta\left(\frac{(\boldsymbol{v}\cdot\boldsymbol{\nabla} T)}{\dot{T}}\right) > W\left(H, C_{\mathrm{p}}, \frac{\partial f_{\mathrm{L}}}{\partial T_{\mathrm{Liq}}}, f_{\mathrm{L}}, \beta\right). \tag{12}$$

For the simple scenario defined by Worster, the thermal gradient is anti-parallel to gravity [8,9,23–25], and there is negligible solidification shrinkage. A pressure gradient arises only due to density changes in the liquid over the depth of the stagnant mushy zone such that Eq. 12 simplifies to:

$$-\frac{\bar{K}g}{\mu}\frac{\Delta\rho_{\mathrm{L}}}{\dot{T}}\frac{\partial T}{\partial z} > f_{\mathrm{L}} W\left(H, C_{\mathrm{p}}, \frac{\partial f_{\mathrm{L}}}{\partial T_{\mathrm{Liq}}}, f_{\mathrm{L}}, \beta = 0\right). \tag{13}$$

Eq. 13 can be re-arranged to obtain the Rayleigh number for directional solidification ($Ra_{\mathrm{DS}}$) proposed by the group of Beckermann for steels [2], Ni-based superalloys and Pb-Sn alloys [1]:

$$Ra_{\mathrm{DS}} = \frac{\Delta\rho_{\mathrm{L}}}{\rho_{\mathrm{L}}^0}\frac{g\bar{K}}{R\nu}. \tag{14}$$

$\nu$ is the kinematic viscosity, $\rho_{\mathrm{L}}^0$ is the density of the original liquid composition at the liquidus temperature, and $R$ is the speed of the solidification isotherms (assuming the relationship $\dot{T} = -R\frac{\partial T}{\partial z}$). The same expression for $Ra_{\mathrm{DS}}$ is obtained using scaling analysis when the characteristic length scale is defined as the ratio of α and $R$ [1,2]. The local-remelting condition is satisfied when $Ra_{\mathrm{DS}}$ exceeds $Ra_{\mathrm{crit}}$, which is given by:

$$Ra_{\mathrm{crit}} = f_{\mathrm{L}} W\left(H, C_{\mathrm{p}}, \frac{\partial f_{\mathrm{L}}}{\partial T_{\mathrm{Liq}}}, f_{\mathrm{L}}, \beta = 0\right). \tag{15}$$

The model does not account for other factors that can also alter or drive fluid flow in the mushy zone, such as specific casting geometries and defects. These are briefly discussed in Section 5.3.

## 4. Results

### *4.1 Determination of $Ra_{\mathrm{crit}}$ for directional solidification of Ni-based superalloys and Pb-Sn alloys*

Single-crystal turbine blades made from Ni-based superalloys are produced using directional solidification [5,40]. These solidification conditions therefore meet very closely the conditions for which Eqs. 14 & 15 can be applied, and data measured by Pollock and Murphy for freckling behavior of the superalloy SX-1 (Table S1.1) is an important dataset for validating our analysis [5]. For these experiments, the casting geometry and alloy composition was identical, and so the results for different cooling conditions are directly comparable. We consider only data for cooling conditions close to the critical conditions for the breakdown of single-crystal growth and the onset of freckling determined experimentally in ref. [5] (Table S1.2).

To determine $Ra$ and $Ra_{\mathrm{crit}}$ across the solidification path, $\Delta\rho_{\mathrm{L}}/\rho_{\mathrm{L}}^{0}$, $\partial f_{\mathrm{L}}/\partial T_{\mathrm{liq}}$, and $v$, $H$ and $C_{\mathrm{p}}$ are calculated as a function of liquid fraction using Thermo-Calc. Methods for calculating $H$ and $C_{\mathrm{p}}$ from Thermo-Calc are provided in Suppl. Mater. S2. For these cooling conditions, it is reasonable to assume there is negligible diffusion in the solid on the length scale of interest, and so the classical Scheil-Gulliver model is used to calculate the liquid's composition and properties, and temperature profiles during solidification. We calculate the primary dendrite-arm spacing $\lambda_1$ using the relation determined in ref. [5] ($\lambda_1 \propto G^{-\frac{1}{2}} R^{-\frac{1}{4}}$) which matches well with classical models and other experimental results [41,42]. The mean permeability of the mush can then be calculated using the Poirier relation [1]:

$$\overline{K} = 0.074\lambda_1^2(-\ln(1 - f_{\mathrm{L}}) - 2f_{\mathrm{L}} - 0.5(1 - f_{\mathrm{L}})^2 + 0.51). \tag{16}$$

The value of $Ra_{\mathrm{crit}}$ for SX-1 monotonically decreases with increasing solid fraction (Figure 3). $Ra$ curves as a function of solid fraction for various studied combinations of $G$ and $R$ reported in ref. [5] are overlaid on the same figure, and they are labelled according to the observed freckling outcome reported by Pollock and Murphy. At low solid fractions, *Ra* is low due to limited solute redistribution which keeps $\Delta\rho_L/\rho_L^0$ low. As the solid fraction increases, the value of *Ra* goes through a maximum due to increasing value of $\Delta\rho_L/\rho_L^0$ but the opposing reduction in $\bar{K}$ with solid fraction. As the cooling rate is reduced, $|\boldsymbol{R}|$ typically reduces, $\lambda_1$ increases and so the value of *Ra* increases for a given alloy composition across the entire solidification path ($0 \leq f_L \leq 1$).

In reasonable agreement with experimental results, freckles are found in the as-cast microstructure for solidification conditions that satisfy the local remelting criterion ($Ra_{DS}$>$Ra_{crit}$). There are some instances where freckling was not found despite *Ra* exceeding $Ra_{crit}$. The error in *Ra* due to uncertainty in $|\boldsymbol{R}|$ and $\lambda_1$ is approximately 20% [1], and so it is within experimental uncertainty that no freckles are observed for some cooling conditions where *Ra* marginally exceeds $Ra_{crit}$ (Case 14). For other cases (Cases 3,4 and 21), *Ra* still substantially exceeds $Ra_{crit,}$ but no freckling was observed (Figure 3). These cases are briefly discussed in Section 5.3.

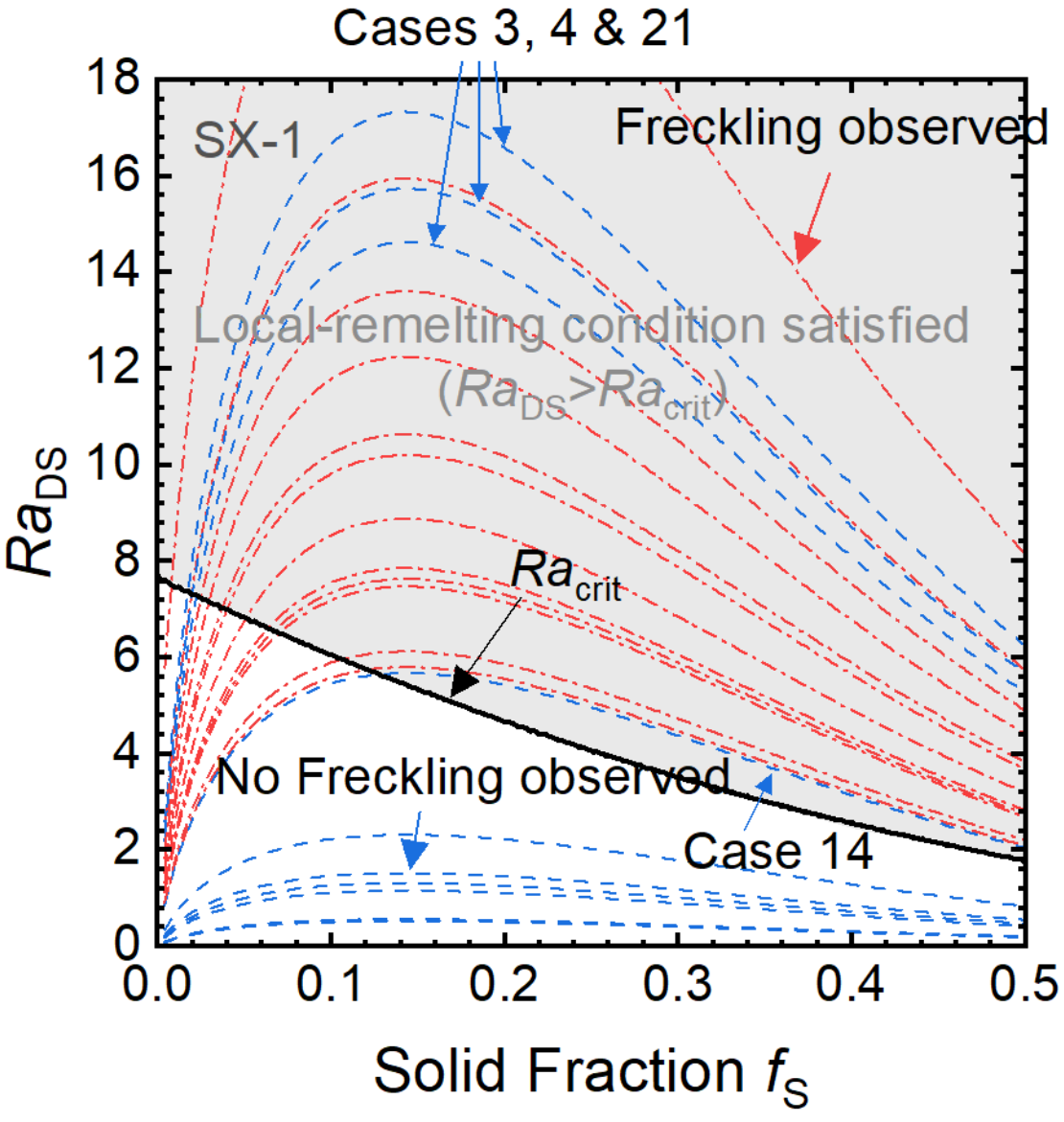

**Figure 3.** *Ra* curves for solidification of the Ni-based superalloy SX-1 under different cooling conditions with the same casting geometry, and the $Ra_{crit}$ curve calculated using CALPHAD. Data for *R* and $\partial T/\partial z$ are taken from the study of Pollock and Murphy [5], which are reported in ref. [1] and reproduced in Table S1.2. The solid black line corresponds to the $Ra_{crit}$ curve computed using CALPHAD. Dashed Blue lines correspond to cooling conditions where no freckling was observed. Red dash-dot lines correspond to conditions where freckling was observed. The shaded region indicates where the local remelting condition is satisfied.

Given the uncertainty in experiments and CALPHAD data, we choose to also evaluate our model by computing *Ra* and $Ra_{crit}$ curves for microscale simulations on directional solidification Pb–Sn binary alloys (Figure 4) [32]. Freckling behavior in Pb–Sn alloys during directional solidification has also been widely studied and empirically-determined values of $Ra_{crit}$ have been reported [1,32].

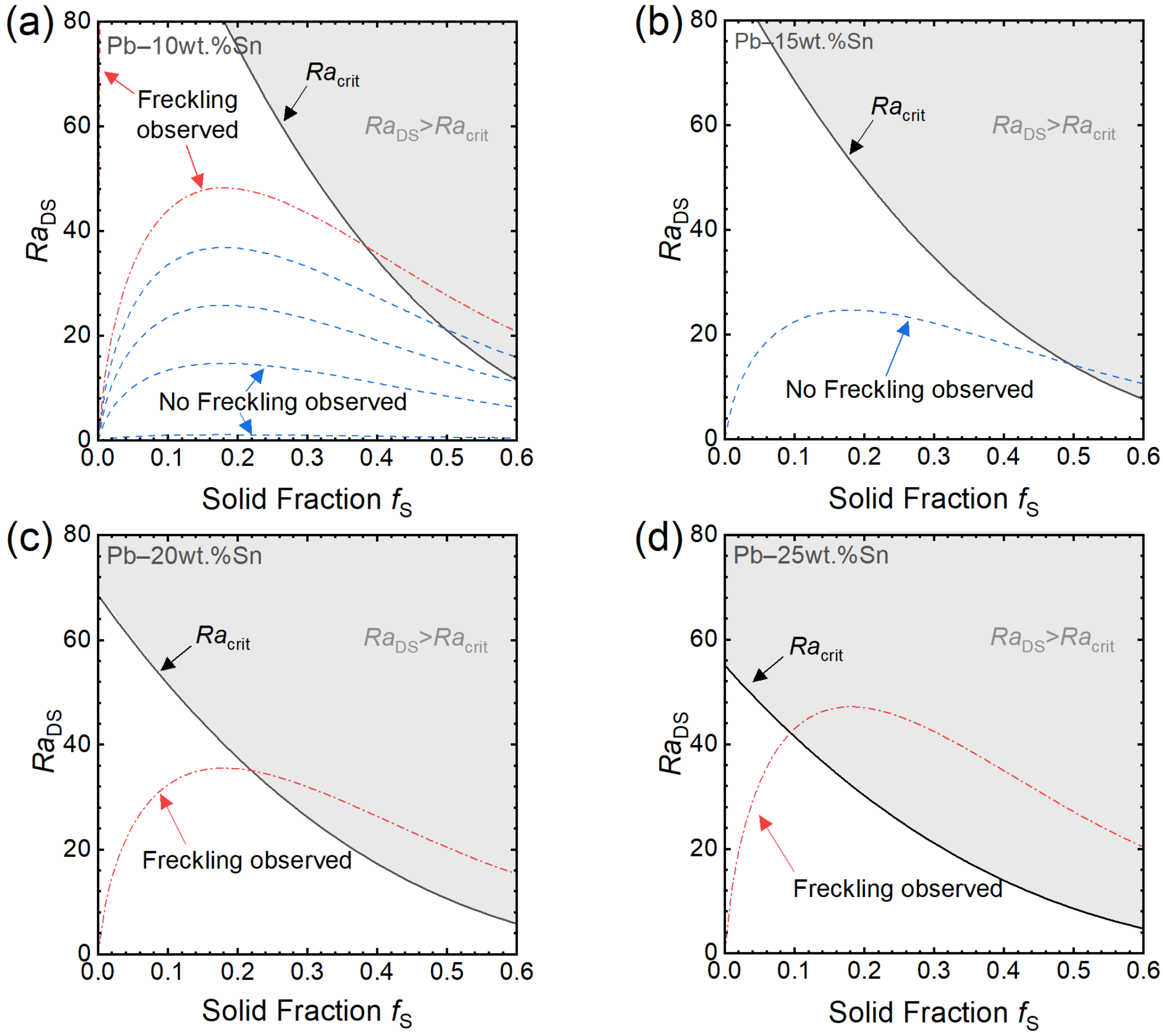

**Figure 4.** *Ra* and $Ra_{crit}$ curves for (a) Pb–10wt.% Sn, (b) Pb–15wt.% Sn, (c) Pb-20wt.% Sn and (d) Pb-25wt.% Sn studied by Yuan and Lee using microscale simulations [32]. These curves are calculated using the cooling conditions and thermophysical properties reported in their study (Tables S3.1 and S3.2). The *Ra* curves are color coded according to the outcome of each simulation. Plots are shown over the first 60% of solidification (where the onset of freckling typically occurs).

For these microscale simulations performed by Yuan and Lee [32], $|\boldsymbol{R}|$, $G$ ($= \partial T/\partial z$) and thermophysical properties are precisely known (Table S3.1 & S3.2), such that these simulations can better evaluate our model. Our predictions for the solidification conditions under which the local remelting criterion is satisfied ($Ra_{DS}$>$Ra_{crit}$) and channels formed during these simulations are in good agreement. There are no instances where *Ra* stays below $Ra_{crit}$, but a channel formed. In some conditions, *Ra* exceeds $Ra_{crit}$ but a channel did not form. This occurs only when the curves intersect at high solid fractions (where the magnitude of *Ra* is small).

By comparison, we also note that the value of $Ra_{crit}$ for Pb–10 wt.% Sn is much larger than that calculated for SX–1; $Ra_{crit}$ effectively scales with $H/C_p$ and the value of the quotient used in these simulations for Pb-Sn alloys is much larger. The values of both *Ra* and $Ra_{crit}$ change noticeably across the Pb-Sn binary composition space due to the variation in density inversion and $\partial f_L / \partial T_{liq}$ with composition (Figure 4).

*4.2 Determination of $Ra_{eff}$ and $Ra_{crit}$ for steel ingot castings*

While single-crystal turbine blades made from Ni-based superalloys are produced using directional solidification, large steel ingots are solidified without such directional control. Fluid flow within an ingot casting is more complex and vigorous flow is expected in the fully liquid region of the casting. Although analytical descriptions of flow patterns, temperature and pressure gradients across the entire casting are not possible, flow in the fully liquid region is suggested to have little effect on the onset of convection in a stagnant mushy zone [9, 43]. The mushy-zone depth is much larger than the dendrite-arm spacing, and so the dynamic influence of the fully liquid region subsequently reduces to a constant pressure on the mush-liquid interface [43].

To explore this application of our model to this more complex scenario, we modify Eqs. 12–15 to describe the stability of an initially stagnant mushy zone which is advancing at an arbitrary orientation with respect to gravity (see Suppl. Mater. S4). We do not consider more complex geometrical effects that can locally accelerate flow [28] or the effects of side-wall losses and imperfect heat transfer [26], since our focus is the effect of composition. We find that $Ra_{crit}$ at a given solid fraction is independent of orientation, but an effective Rayleigh number $Ra_{eff}$ varies with the orientation of the solidification front with respect to gravity such that $Ra_{eff}(\theta) = |Ra_{DS}| \cos\theta$. $\theta$ is the angle of $\nabla T$ relative to $\boldsymbol{g}$.

In previous work, Torabi Rad *et al.* determined a $Ra_{crit}$ value for steel ingots by approximating the scenario to that of vertical directional solidification and using empirical methods to determine *Ra* at which the onset of A-type segregation was reported [2]. They used data from Yamada *et al.* and Suzuki & Miyamoto for which the susceptibility to A-type segregation was studied in both large commercial ingots (9–400 tonnes) and using a small experimental setup (15 kg ingots) designed to simulate these larger ingots by studying unidirectional solidification when cooled from the side [12–14,45]. Suzuki and Miyamoto found that when $R^{2.1}G$ fell below a critical value $S_{crit}$, A-type segregates formed in these ingots. The value of $S_{crit}$ is particular to each alloy composition, with a higher value of $S_{crit}$ corresponding to a composition which is more susceptible to A-type segregate formation.

In the present work, values of $S_{crit}$ for these twenty-seven experimental ingots (Table S5.1) are compared to our CALPHAD calculations of $Ra_{eff}$ and $Ra_{crit}$ using Pandat's PanIron Database. To compute $Ra_{eff}$ and $Ra_{crit}$, it is necessary to make a series of assumptions about local solidification conditions in the ingot. For the small experimental ingots performed in refs. [14,15], it was found that A-type segregates were inclined by an angle $\theta_A$ between 0–45° from the vertical in the as-cast microstructure according to the relationship [14,15]:

$$\theta_A(^\circ) = 0.66R. \tag{17}$$

This differs from the inclination of the solidification front, which Suzuki and Miyamoto found $\boldsymbol{R}$ to lie at approximately 80° to the vertical in this setup [15,16]. Process simulations of large ingots suggest that $\nabla T$ and $\boldsymbol{R}$ lie at an angle $\theta$ of 75–80° about the vertical for a typical steel ingot [44]. The small-scale experiments therefore appear to be a suitable proxy for large ingot castings. We therefore choose to consider $Ra_{eff}$ at effective tilt angles of 0° and 80° for computing $Ra_{eff}$ in this work.

In contrast to the assumptions made in ref. [2], we do not assume a fixed reference density at $T^0_{liq}$. CALPHAD calculations suggest that the variation in $\rho_L^0$ (Figure 5b) is an order of magnitude larger than $\Delta\rho_L$ over the initial portion of the mush (Figure 5c). This variation in $\rho_L^0$ scales with $T^0_{liq}$, suggesting there is a substantial effect from the variation in liquidus temperature between alloys. Calculations suggest $\Delta\rho_L$ has both positive and negative values across the twenty-seven alloy compositions. Numerical simulations suggest the horizontal-component of solidification enables both directions of density change to cause channel formation: positive buoyancy leads to 'upward oriented' A-type segregates; negative buoyancy leads to 'downward oriented' A-type segregates (Figure 5c). In the latter case, A-type segregates would form in regions of the mushy zone where there is outward flow in response to the negative buoyancy inversion [44].

As in ref. [2], $Ra$ and $Ra_{crit}$ are determined at a single value of solid fraction (30%). This is based on the undercooling at which the horizontal components of $\boldsymbol{R}$ and $\boldsymbol{G}$ were measured and A-type segregation occurred. Yamada *et al.* suggested this 15 K undercooling corresponds to 30–35% solid fraction for these alloys [45], which is in good agreement with CALPHAD calculations (12±5 K at 30% solid fraction). $|\boldsymbol{R}|$ is estimated to be 55±30 μm s$^{-1}$ for "small experimental ingots" and 28±10 μm s$^{-1}$ for "large commercial ingots" [2]. We assume these measurement horizontal components are equal to $|\boldsymbol{R}|$ and $|\boldsymbol{G}|$—an assumption that appears reasonable given the solidification front is inclined at ~80˚.

The permeability of the local average volume element for steel ingots was calculated using Eq. 16. Although it is frequently used, the Blake-Kozeny relation has been suggested inappropriate for this definition of $Ra$ [1]. Without experimental data to inform a more accurate calculation, we assume, as in ref. [2], that $\lambda_1$ is twice the secondary dendrite arm spacing ($\lambda_2$). $\lambda_2$ is calculated using the two expressions suggested in ref. [2] (see Suppl. Mater. S5). Of these expressions, we find neither relationship provides an excellent fit to the values measured by Yamada *et al.* [45] (Figure 5a). This is partly due to substantial uncertainty inherited from estimates of $|R|$, but both expressions appear to exaggerate the trend that $\lambda_2$ decreases with increasing $S_{crit}$ [45].

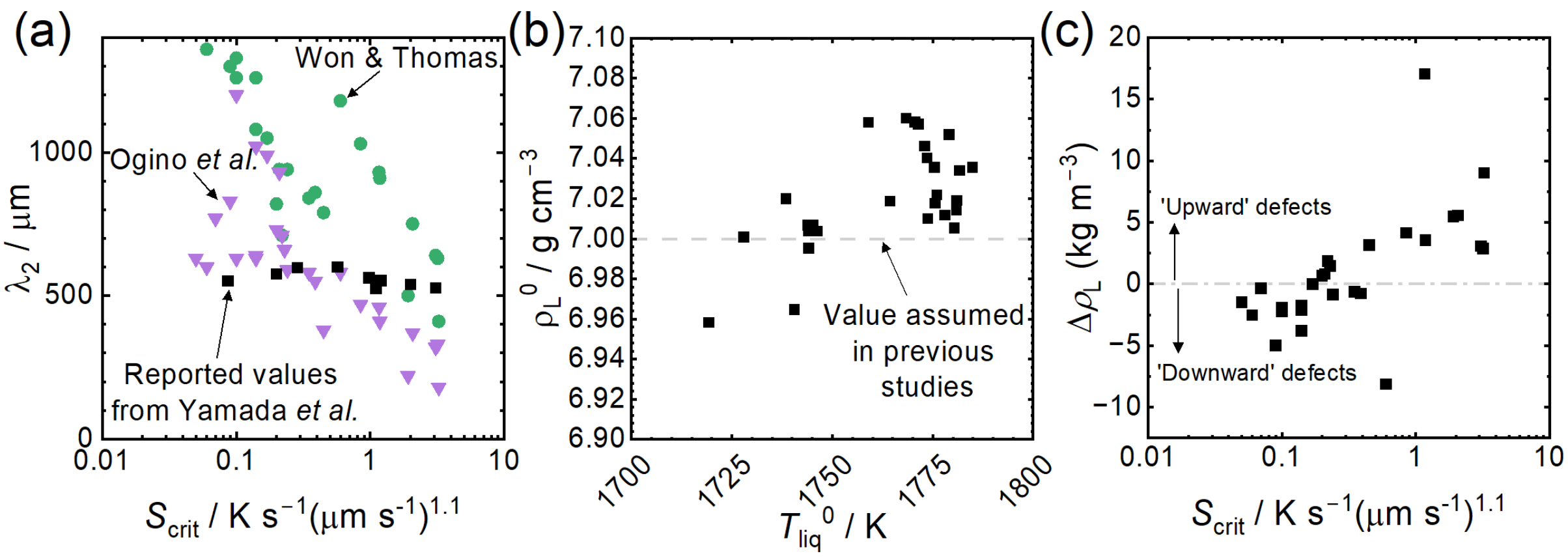

**Figure 5.** (a) Plot of $\lambda_2$ and $S_{crit}$ using the two empirical relations considered in ref. [2] for the twenty-seven steel ingots. Shaded region highlights the range in which values were empirically measured for a select number of ingots by Yamada *et al.* (b) Plot of the calculated liquidus temperature of the alloy $T^0_{liq}$ against the density of the liquid at this temperature. The fixed density value of 7.002 g cm$^{-3}$ assumed by Torabi Rad *et al.* is also shown for comparison. (c) Plot of calculated $\Delta\rho_L$ against $S_{crit}$ for these compositions.

Figure 6a shows $Ra_{crit}$ against $S_{crit}$ at 30% solid fraction. This suggests that $Ra_{crit}$ decreases with increasing $S_{crit}$, such that compositions with lower $Ra_{crit}$ are more susceptible to A-type segregation.

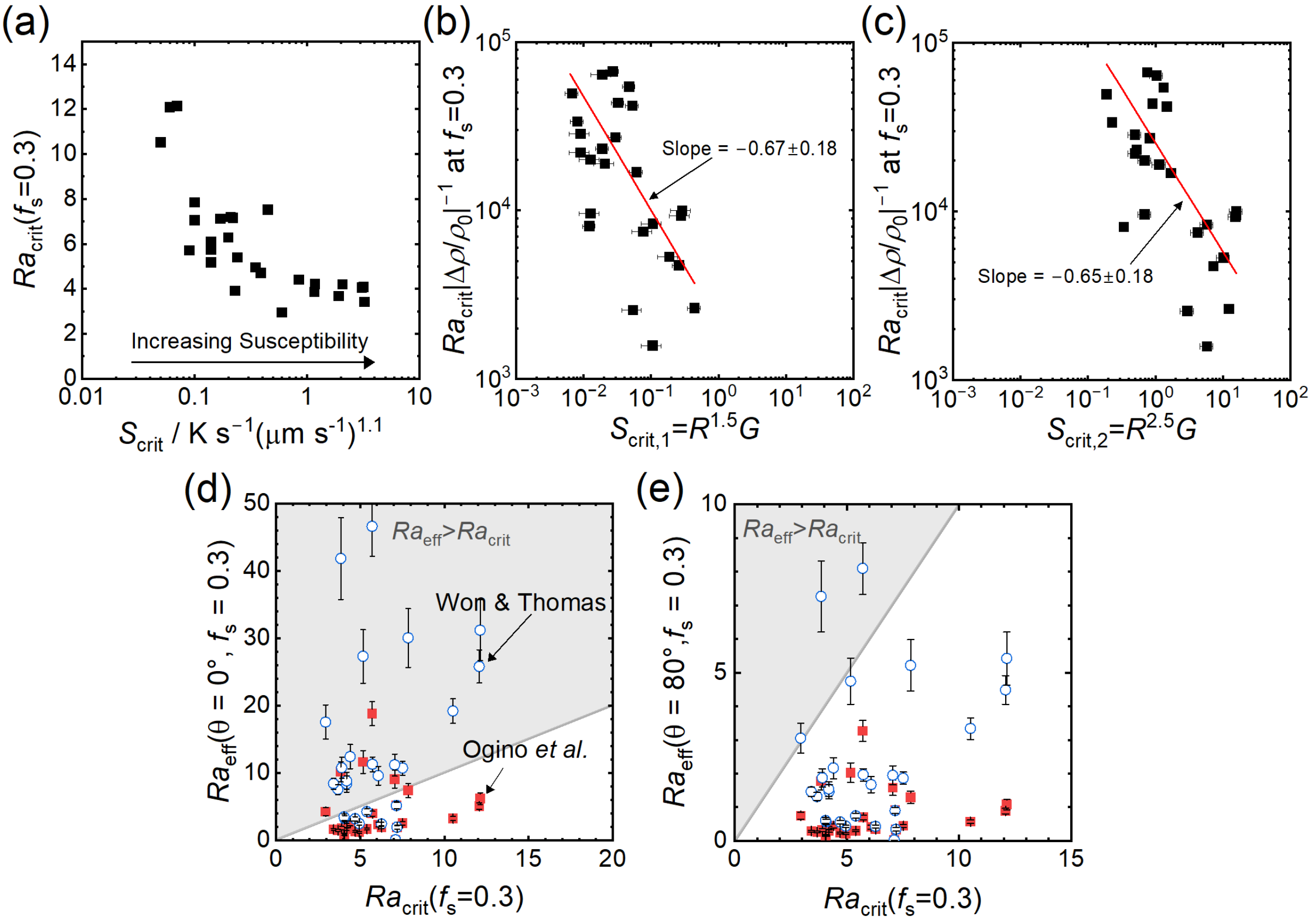

**Figure 6.** (a) Plot of $Ra_{crit}$ at 30% solid fraction ($f_s$) against $S_{crit}$ for the twenty-seven steel ingots studied in ref. [13–15]. (b) Plot of $Ra_{crit}|\Delta\rho/\rho_0|^{-1}$ against $S_{crit,1}$. (c) Plot of $Ra_{crit}|\Delta\rho/\rho_0|^{-1}$ against $S_{crit,2}$. (d) Plot of calculated $Ra$ against $Ra_{crit}$ at $f_S$=30% for 0˚ effective tilt. Data used to compute $Ra$ and $Ra_{crit}$ is given in Table S5.1. (e) same ingots as (d) but for $Ra$ plotted at 80˚ effective tilt. For these compositions, the kinematic viscosity was assumed to be $8\text{x}10^{-7}$ m $s^{-1}$ [2]. The error bars correspond to the uncertainty propagating from the value's dependence on $|\boldsymbol{R}|$ using uncertainty estimates reported in ref. [2]. This is a lower-bound estimate of the overall uncertainty in $Ra_{eff}$.

According to our model, $Ra_{eff}$ should equal $Ra_{crit}$ when the local-remelting condition is first satisfied. If $Ra_{eff}$ scales with $\lambda_2^2$ (where $\lambda_2 \propto (GR)^{-1/3}$ ), then $Ra_{eff} \propto (R^{2.5}G^1)^{-\frac{2}{3}}$. Assuming $Ra_{crit}|\Delta\rho_L/\rho_0|^{-1}$ contains most composition effects, then $Ra_{crit}|\Delta\rho_L/\rho_0|^{-1}$ is proportional to $S_{crit,2}^{-2/3}$ where $S_{crit,2}$ represents a critical combination of $R^{2.5}G^1$. Alternatively,

if $Ra_{\text{eff}}$ scales with $\lambda_1^2$ (where $\lambda_1 \propto G^{-\frac{1}{2}}R^{-\frac{1}{4}}$) [42,46], then $Ra_{\text{crit}}|\Delta\rho_L/\rho_0|^{-1}$ is expected to scale with $(S_{\text{crit},1})^{-1}$, where $S_{\text{crit},1}$ represents a critical value of $G^1R^{1.5}$.

Figures 6b and 6c compare the computed values of $Ra_{\text{crit}}|\Delta\rho/\rho_0|^{-1}$ with the suggested expressions $S_{\text{crit},1}$ and $S_{\text{crit},2}$. Linear regression fits suggest plausible agreement with the power-law dependencies given the uncertainties in these estimates. The original empirical expression for $S_{\text{crit}}$ has power-law dependencies on $R$ and $G$ that fall between these two models, suggesting reasonable agreement between our model and experimental results.

Plotting $Ra_{\text{crit}}$ against $Ra_{\text{eff}}$ at 0° effective tilt (Figure 6d), $Ra_{\text{eff}}(f_s=0.3)$ typically exceeds $Ra_{\text{crit}}(f_s=0.3)$, which would suggest the local remelting condition is satisfied at solid fractions below 30%. By contrast, assuming an 80° effective tilt (Figure 6e), the values of $Ra_{\text{eff}}(f_s=0.3)$ are typically lower than $Ra_{\text{crit}}$. This implies the onset of channel formation may not occur by local remelting, or only at higher solid fractions. Since the uncertainty in $|\boldsymbol{R}|$ is estimated to be as high as 50%, and the predicted values of $\lambda_2$ differ by as much as a factor of 2 from experimental measurements for a select number of alloys (Figure 5a), an exact explanation is unclear from the present work. Although the suggested description of compositional effects from this model appears broadly consistent with experimental results, assessing the ability of this local remelting criterion to predict A-type segregation in large ingots requires more precise and accurate knowledge of local solidification conditions to construct a representative volume element, and a more rigorous validation of thermophysical properties.

## **5. Discussion**

### 5.1 Physical meaning of $Ra$ and $Ra_{\text{crit}}$

Mehrabian *et al.* classified the susceptibility to channel formation in three different regimes: "stable", "intermediate", and "unstable" flow [6]. In the stable regime, $\boldsymbol{v} \cdot \boldsymbol{\nabla} T < 0$. Hot liquid of bulk composition flows into the mushy zone, primarily to account for solidification shrinkage. It is not possible for local remelting to occur, and channel formation is not expected in an isotropic and uniformly varying mushy zone. For $\boldsymbol{v} \cdot \boldsymbol{\nabla} T = 0$, liquid in the mushy zone is stagnant. In the absence of other effects, and if the liquid is neutrally buoyant—that is, there is no change in the density of the liquid during solidification—the liquid would remain stagnant as further solidification occurs (Figure 7a). Channels are not expected to form [48], and the residual interdendritic liquid will eventually solidify with a near-eutectic composition, or otherwise, such that only micro-segregation is observed.

When $\boldsymbol{v} \cdot \nabla T > 0$, flow can either be "intermediate" or "unstable". Within Mehrabian *et al.*'s description [6], buoyancy inversion across a stagnant mushy zone drives outward flow of solute-rich liquid from the solid-mush interface. A small liquid parcel would begin to flow outward through the mushy zone. This liquid parcel is no longer in thermodynamic equilibrium with the nearby solid and to re-establish equilibrium, it remelts some of the solid. The outward flow reduces the solidification rate (equal to $-\partial f_L/\partial t$) and the magnitude of the impulse to initiate local remelting and the onset of channel formation is reduced. In the "intermediate" regime, there is sub-critical outward flow such that local remelting cannot occur, but the solidification rate is suppressed (Figure 7b,c).

For scenarios where $\boldsymbol{v} \cdot \nabla T$ already satisfies Eq. 9, the flow is "unstable" and local remelting will occur. Cold solute-rich liquid redissolves hot solid regions of the mushy zone to form channels that cannot re-solidify. At the onset of local remelting, $\partial f_L/\partial t$ is equal to zero and so the function $W$ is equal to 1, reducing to the original criterion proposed by Flemings.

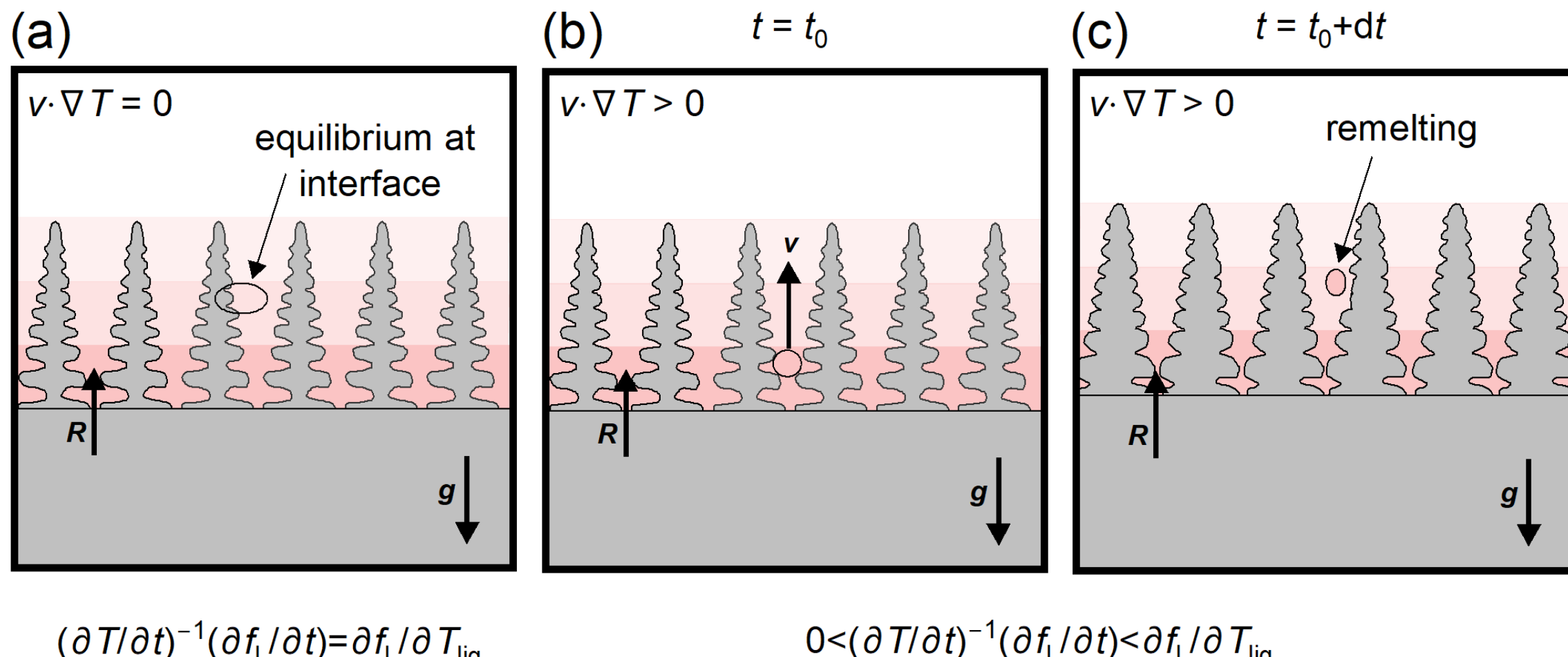

**Figure 7.** (a) Schematic of scenario where $\boldsymbol{v} \cdot \nabla T = 0$, and the density of the residual liquid does not change over the solidification path. In the stagnant mushy zone, the liquid and the solid at the interface between them remain in thermodynamic equilibrium. (b) $\boldsymbol{v} \cdot \nabla T > 0$ and so there is outward flow of solute-rich liquid from the interface. (c) At a later time $t_0$+d$t$ solute-rich liquid that advances upwards begins to remelt part of the mushy zone to re-establish local thermodynamic equilibrium with the nearby solid. This results in a lower solid fraction and dilutes the solute-rich liquid. The solidification rate is now lower than that expected for a stagnant mushy zone.

Within a $Ra$ number description, it is well established that $Ra$ describes the ratio of the buoyancy forces to the resistance to flow by the porous mushy zone [8]. In a formal stability analysis, $Ra_{\mathrm{crit}}^{W}$ defines the onset of natural convection due to a loss of linear stability in the non-convecting mushy-layer base state (stagnant mushy zone). Our expression for $Ra_{\mathrm{crit}}$ derived using Flemings' heuristic model (Eq. 15) instead describes the magnitude of the buoyancy-driven flow, when suddenly triggered, required to arrest solidification of an initially stagnant mushy zone. It effectively delineates the boundary between the "unstable" and "intermediate" regimes proposed by Mehrabian *et al.* [6]. Within an infinitesimal time interval d$t$, the buoyancy-driven flow must be fast enough to melt the solid fraction d$f_{\mathrm{L}}$ that would have otherwise formed in the time d$t$ and melt enough solid to lower the temperature by an amount d$T$ (Figure 7b,c). When the fluid flow is sufficient to advect heat and solute above a critical rate, outer regions of the mushy zone are effectively cooled by local remelting rather than by thermal conduction. The significance of the evolution of latent heat on the stability of the mushy zone has been previously highlighted by Amberg & Homsy, and Anderson & Worster [26,28], but is not evident from empirical evaluations of $Ra_{\mathrm{crit}}$ for metal alloys and its correlation with channel formation.

## 5.2 Composition dependence of $Ra$ and $Ra_{\mathrm{crit}}$ and implications for alloy design

Our re-expression of Flemings' model in a Rayleigh-number form suggests $Ra_{\mathrm{crit}}$ for local remelting not only varies between alloy families, but it has a strong dependence on composition within a relatively narrow range. The composition dependence of $Ra$ and $Ra_{\mathrm{crit}}$ is therefore important for the design of casting alloys that are resistant to freckles and A-type segregate formation, or when choosing between existing alloys for a particular casting. The compatibility of our expressions with CALPHAD offers an opportunity to integrate $Ra$ and $Ra_{\mathrm{crit}}$ as parametric design constraints and selection criteria within a systems-design approach to alloy development [16,17].

At a given solid fraction, $Ra_{\mathrm{crit}}$ is a function of $H$, $C_{\mathrm{p}}$, and $\partial f_{\mathrm{L}} / \partial T_{\mathrm{liq}}$, and their combination represents an apparent Stefan number $Ste_{\mathrm{app}}$ which is a function of the local average solid fraction:

$$\left[Ste_{\mathrm{app}}(f_{\mathrm{L}})\right]^{-1} = \frac{H}{C_{\mathrm{p}}} \frac{\partial f_{\mathrm{L}}}{\partial T_{\mathrm{liq}}}, \tag{18}$$

such that:

$$Ra_{\mathrm{crit}} = f_{\mathrm{L}}\left(1 + \left[Ste_{\mathrm{app}}(f_{\mathrm{L}})\right]^{-1}\right). \tag{19}$$

Compositions with a low $Ste_{\mathrm{app}}$ would therefore appear generally more resistant to local remelting during solidification. For compositions where $\Delta\rho_{\mathrm{L}}$ is also minimized across the solidification path and $\overline{K}$ is small, slow solidification is required for $Ra$ to exceed $Ra_{\mathrm{crit}}$ — these alloy compositions are most resistant to freckle formation. In the simple case of binary Pb-Sn alloys, we find that increasing the Sn content from 10 wt.% to 25 wt.% has two major effects: it reduces $Ra_{\mathrm{crit}}$ by almost 60% due to solute partitioning of Sn reducing the $\partial f_{\mathrm{L}} / \partial T_{\mathrm{liq}}$; it increases $Ra$ by a factor of 2.5 due to the effect of Sn enrichment on buoyancy inversion. Their combined effect is to restrict the range of solidification conditions where the local remelting condition is not satisfied — the critical value of $R^{-1.5}G^{-1}$ decreases by a factor of 6.

For the range of steel alloys studied, we show that $Ra_{\mathrm{crit}}$ correlates inversely with $S_{\mathrm{crit}}$. This suggests compositions that have high $H$, low $C_{\mathrm{p}}$ and have a narrow freezing range (high $\partial f_{\mathrm{L}}/\partial T_{\mathrm{liq}}$) are typically most resistant to A-type segregation. Our conclusions agree well with experimental findings by Yamada *et al.* [45] who found that steels which (a) solidify in a narrow temperature range, and (b) have minimal buoyancy inversion across the mushy zone, have lower values of $S_{\mathrm{crit}}$ (Figure S6.1).

## 5.3 Considerations beyond the present model

Our re-expression of Flemings' model in Rayleigh-number form considers an idealized description of the mushy zone. This enables us to consider compositional effects on the susceptibility to channel formation in metal castings, and derive practical expressions for $Ra$ and $Ra_{\mathrm{crit}}$ by conserving mass, solute and energy within the local average volume element. Still, it is important to discuss the limitations of our model and briefly consider the significance of results from recent numerical studies on channel formation in metal castings.

### *5.3.1 Additional Modes of Convection*

Several full linear stability analyses find that the direct (non-oscillatory) mode of convection is typically bimodal [22,49,50]; a boundary-layer mode of convection that occurs within liquid just above the mush-liquid interface (the compositional boundary layer seen in Figure 2), and a mushy-layer mode of convection which is the focus of the present work. The former results in fine-scale double-diffusive convection in the fully liquid region directly above the mushy zone. Although this boundary-layer convection mode is established long before discernible fluid flow through the mushy zone, such fingering has little effect on when the onset of convection occurs [49]. In the present model, the compositional boundary layer is effectively suppressed by the assumptions made. Our model forces the Lewis number to infinity [27], which is a reasonable approximation for metallic systems [32], but may be less appropriate for aqueous ionic solutions. Within the mushy zone itself, the buoyancy field is dominated by solutal effects such that no double-diffusive interactions between the solute and thermal fields occur [9]. The good agreement of the present analysis with experimental results appears consistent with these arguments and helps to justify the simplifications of complex interactions in mushy zones made in the present work.

### *5.3.2 The formation of channels when the local-remelting condition is not satisfied*

Linear stability analyses based on models assuming a planar solidification front consider only the stability of the mushy zone in response to infinitesimal perturbations [47,48]. Marginal stability analysis and weakly non-linear stability analyses under near-eutectic and large far-field temperature asymptotic limits have highlighted further modes of convection that may occur when $Ra<Ra^{W}_{\mathrm{crit}}$ (subcritical bifurcation) [22–24]. These modes, which may also lead to channel formation occur in response to perturbations of finite amplitude [24]. This mechanism is not considered in the present work.

Numerical simulations of metal castings have suggested the effects of defects, second phases, inclusions, and dendritic overgrowth on influencing channel formation [9,18,19,32]. Studies suggest that they can lead to channel formation, without satisfying the local remelting criterion. Kumar *et al.* considered the effect of a small perturbation to the local-average value of $f_L$ within Flemings' volume element [50]. Using a linear perturbation analysis of the Darcy equation, they suggest that, in the infinitesimal limit, perturbations of the liquid fraction can be readily amplified leading to the formation of channels [50]. In a different approach but using a similar ideal mushy zone to the present work, Simpson *et al.* considered local remelting about a well-developed needle-like defect in the mushy zone [10]. The presence of such defects can substantially reduce the local value of $\overline{K}$, accelerating local flow and thereby increasing the susceptibility to freckling. They suggest the effect is so strong that freckling may conceivably occur in the "stable" flow regime [10].

The coupled, non-linear behavior of the flow, composition and temperature fields are not considered in this work. As such, the present work does not consider behavior leading to channel formation in Mehrabian *et al.*'s intermediate regime. Extensions of the model to provide appropriate descriptions for these effects are important.

*5.3.3 Channel evolution and instabilities in the local solidification rate*

In microscale simulations of Pb–Sn alloys, Yuan and Lee found that solute-rich liquid trapped at the dendrite tips may suppress the solidification rate [32]. Whether a channel forms and is sustained, or not, is not only due to outward flow but also "a complex balance between secondary dendrite arm remelting and primary arm deflection and overgrowth" [32]. They proposed that the evolution of a channel when $Ra > Ra_{\mathrm{crit}}$ is therefore, to some extent, "stochastic" [32].

As discussed by Worster [9], an $Ra$-based stability analysis describes the onset of natural convection in a non-convecting base state (stagnant mushy zone). It does not describe the subsequent evolution of a channel from the perturbation. Once convection begins, compositional equilibrium is quickly lost, and non-linear terms are required to accurately describe whether a channel develops and how the liquid flows within it [9]. The present model considers only whether a buoyancy-driven flow would be sufficient to satisfy the local-remelting condition. It does not describe whether a channel will ultimately form.

In Figure 3 and 4, there are therefore examples where $Ra$ exceeds $Ra_{\mathrm{crit}}$ and the local remelting condition is satisfied, but sustained channels that lead to freckles are not found in experiments or numerical simulations. For the Pb-Sn simulations in Figure 4, these examples are found only where $Ra$ exceeds $Ra_{\mathrm{crit}}$ at high solid fractions; where the low permeability of the mush inhibits fluid flow. In these cases, it might be expected that the growth of perturbations to the planar front into channels is slow. For cases 3, 4 and 21 for solidification of SX-1 (Figure 3), freckles are not found despite the local remelting condition being satisfied at low solid fractions. Solidification occurs under conditions where $|\boldsymbol{G}|$ is high and $|\boldsymbol{R}|$ is low compared to other cases studied (see Table S1.2); the solidification front advances slowly and the mushy zone is comparably thin. It is unclear from the present analysis whether the assumptions for this model hold, or whether these conditions inhibit the growth of perturbations into channels. Extending the present model to understand the evolution and the role of other effects on channel formation, and complementing it with numerical simulations may help to develop alloys and more advanced casting methods in which freckling and A-type segregation are consistently not found.

## 6. Conclusions

Expressions for the Rayleigh number $Ra$ were derived by considering Flemings' model for an idealized, initially stagnant mushy zone. A critical value of $Ra$, denoted $Ra_{crit}$, was derived above which the local-remelting condition is satisfied. Previous reports of a single value for $Ra_{crit}$ in metal castings are not supported by the derived expressions; $Ra_{crit}$ is a function of the latent heat of fusion, the change in solid fraction with liquidus temperature and the local specific heat capacity. Since these properties vary even within relatively narrow composition ranges, a single value of $Ra_{crit}$ cannot be assumed for a particular alloy family. In addition, $Ra_{crit}$ varies with local average solid fraction in the mushy zone during solidification, such that $Ra$ and $Ra_{crit}$ should be computed along the entire solidification path.

For the nickel-based superalloy SX-1 and several Pb–Sn alloys, there is good agreement between the local-remelting condition being satisfied and the observation of freckles in selected experimental and numerical studies. For steel ingots, the model shows consistent trends with the empirical critical Suzuki number for the reported alloy compositions. The model helps to explain why steel compositions with narrow freezing ranges and minimal buoyancy inversion are most resistant to A-type segregation.

## Acknowledgments

Part of this research was conducted by UT-Battelle, LLC, under Contract No. DE-AC05-00OR22725 with the U.S. Department of Energy. The authors are grateful to Prof. Merton C. Flemings and Prof. Matthew J. Krane for many useful discussions. They also thank Prof. Matthew J. Krane for critical reading of the manuscript.

## Funding Statement

This research has been funded by the Department of Defense's Industrial Base Analysis and Sustainment (IBAS) Program. Research was conducted under Contract No. DE-AC05-00OR22725 with the U.S. Department of Energy.

## Conflict of Interest Statement

The authors have no conflicts of interest to disclose.

# Composition dependence of the critical Rayleigh number curve for macrosegregation in multicomponent metal alloys

O.S. Houghton, A.S. Sabau, G.B. Olson

## Supplementary Material

### S1. Experimental and Computed data for Ni-based superalloy SX-1

**Table S1.1.** Alloy composition Ni-based superalloy SX-1 from Pollock and Murphy [5].

| Ni | Al | Cr | Co | Hf | Re | Ta | W |
|---|---|---|---|---|---|---|---|
| Bal. | 6.0 | 4.5 | 12.5 | 0.16 | 6.3 | 7.0 | 5.8 |

**Table S1.2.** Experimental data for vertical, unidirectional solidification of the Ni-based superalloy SX-1 from Pollock and Murphy [5]. $\lambda_1$is calculated using the expression $\lambda_1 = 5000\ \mathrm{G}^{-1/2}\ R^{-1/4}$, as determined from Figure 3 in ref. [5].

| Experiment | $R$ (μm s$^{-1}$) | $G$ (K cm$^{-1}$) | $\lambda_1$ (μm) | Freckles reported |
|---|---|---|---|---|
| 3 | 14.00 | 96.00 | 264 | no |
| 4 | 14.00 | 81.00 | 287 | no |
| 5 | 56.00 | 23.00 | 381 | yes |
| 6 | 56.00 | 16.50 | 500 | yes |
| 7 | 56.00 | 12.90 | 509 | yes |
| 8 | 56.00 | 11.00 | 551 | yes |
| 9 | 56.00 | 7.70 | 659 | yes |
| 10 | 70.00 | 96.10 | 176 | no |
| 11 | 100.00 | 31.80 | 280 | no |
| 13 | 110.00 | 11.00 | 466 | yes |
| 14 | 113.00 | 10.80 | 467 | no |
| 15 | 113.00 | 10.00 | 485 | yes |
| 16 | 113.00 | 8.20 | 536 | yes |
| 18 | 176.00 | 60.50 | 176 | no |
| 19 | 176.00 | 56.20 | 183 | no |
| 21 | 14.00 | 89.20 | 274 | no |
| 27 | 70.00 | 109.00 | 166 | no |
| 28 | 100.00 | 48.50 | 227 | no |
| 29 | 113.00 | 7.80 | 549 | yes |
| 30 | 113.00 | 6.90 | 584 | yes |
| 31 | 113.00 | 6.00 | 626 | yes |
| 32 | 113.00 | 5.00 | 686 | yes |

**S2. Determination of $H$ and $C_p$ for the present model from CALPHAD simulations.**

In CALPHAD simulations, the apparent heat capacity is typically computed in Scheil-Gulliver simulations. This is required for solidification modelling where the temperature field, but not the liquid fraction, is solved. The apparent heat capacity $C_{\mathrm{p}}^{\mathrm{app}}$ is given by [51]:

$$C_{\mathrm{p}}^{\mathrm{app}} = \frac{\mathrm{d}H_{\mathrm{sys}}}{\mathrm{d}T} = f_{\mathrm{L}} C_{\mathrm{p}}^{\mathrm{liq}} + (1 - f_{\mathrm{L}}) C_{\mathrm{p}}^{\mathrm{solid}} + H \frac{df_{\mathrm{L}}}{dT}, \tag{S2.1}$$

where $H_{\mathrm{sys}}$ is the heat of the system (the volume element of interest within the mushy zone), $C_{\mathrm{p}}^{\mathrm{liq}}$ is the specific heat capacity of the liquid, and $C_{\mathrm{p}}^{\mathrm{solid}}$ is the specific heat capacity of the solid. The specific heat capacity of the system $C_{\mathrm{p}}$ is given by:

$$C_{\mathrm{p}} = f_{\mathrm{L}} C_{\mathrm{p}}^{\mathrm{liq}} + (1 - f_{\mathrm{L}}) C_{\mathrm{p}}^{\mathrm{solid}}, \tag{S2.2}$$

For SX-1, the value of $C_{\mathrm{p}}$ computed directly from $C_{\mathrm{p,app}}$ in Thermo-Calc's Scheil-Gulliver model was unphysically large due to numerical errors in $C_{\mathrm{p}}$ for each phase about $T_{\mathrm{liq}}^{0}$. This resulted in a low value of $Ra_{\mathrm{crit}}$ (Figure S2.1). Instead, the value of $C_{\mathrm{p}}$ was calculated using Eq. S2.2. $C_{\mathrm{p}}^{\mathrm{liq}}$ and $C_{\mathrm{p}}^{\mathrm{solid}}$ were calculated using equilibrium calculations. The expected liquid and solid compositions, and temperature for a given local solid fraction were computed using the Scheil-Gulliver model.

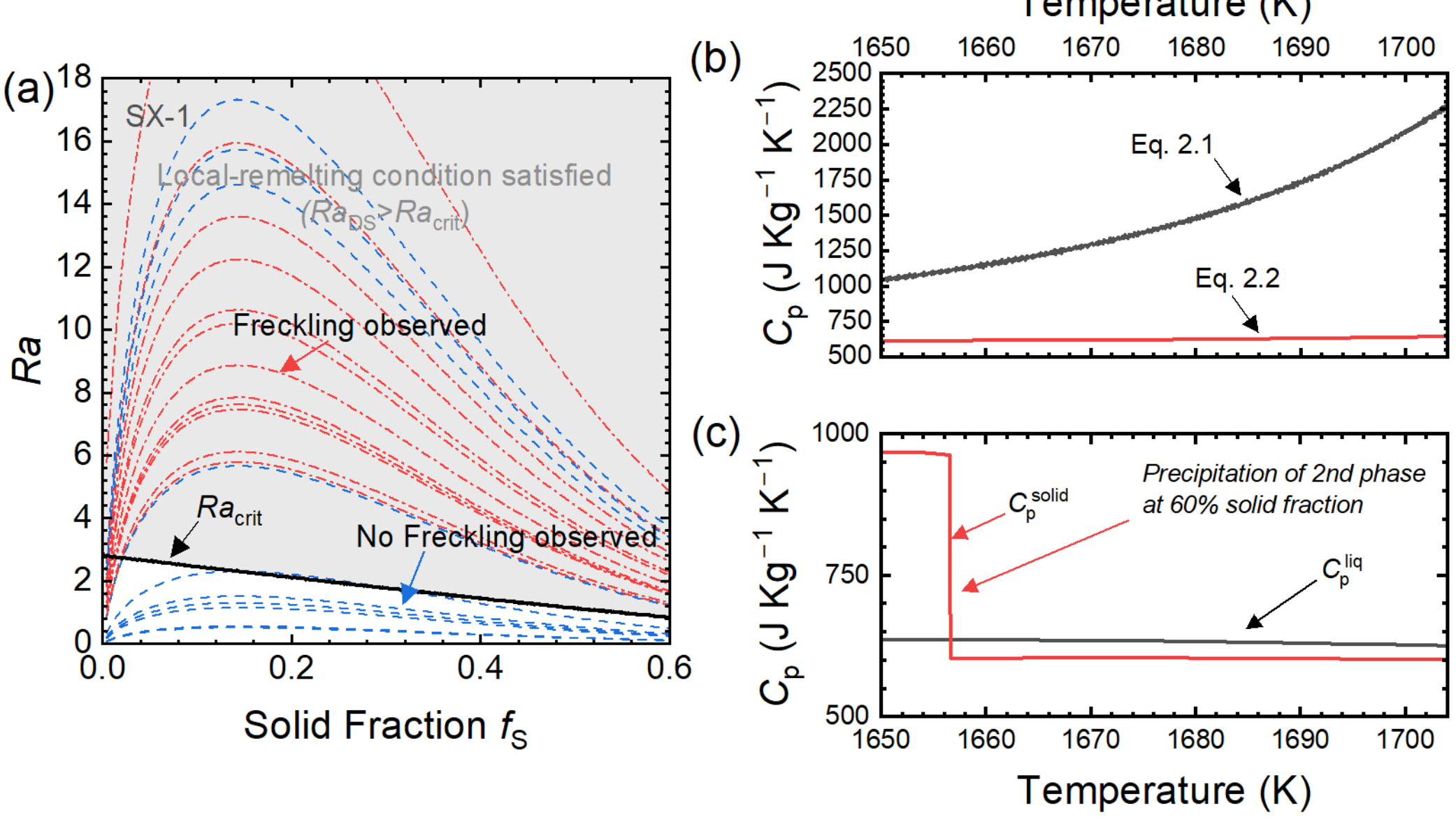

**Figure S2.1.** (a) $Ra$ and $Ra_{\mathrm{crit}}$ curves where $Ra_{\mathrm{crit}}$ is calculated using Eq. 1.1 to determine $C_{\mathrm{p}}$ from the Scheil-Gulliver solidification model. (b) $C_{\mathrm{p}}$ calculated using Eq. 2.1 and Eq. 2.2 over the temperature range of interest. (c) Plot of $C_{\mathrm{p}}$ for solid (ccp) and liquid phases over the temperature range of interest from equilibrium calculations. The arrows indicate the values used to compute $C_{\mathrm{p}}$ using Eq. 2.2.

The latent heat of fusion as a function of solid fraction for SX-1 was calculated using the Scheil-Gulliver calculation. It varied between 190–210 kJ $kg^{-1}$ between the liquidus temperature and 50% solid fraction. These calculated values are consistent with experimental measurements on similar alloys [52].

For evaluation of $Ra_{crit}$ for Pb–Sn alloys, Eq. S2.2 is used to calculate $C_p$ using data provided in ref. [32] (see Table S3.1). For the selected steel compositions, $C_p$ is calculated by computing $H_{sys}$ and $H$ as a function of solid fraction using the Questek's ICMD toolkit, Computherm Software and PanIron 2025 database [37–39]. From the Scheil-Gulliver solidification model, $C_p$ was calculated directly from the solidification data using:

$$C_{\mathrm{p}} = \frac{\mathrm{d}}{\mathrm{d}T}\left(H_{sys} - H f_{\mathrm{L}}\right) = \frac{\mathrm{d}H_{\mathrm{sens}}}{\mathrm{d}T}, \tag{S2.3}$$

where $H_{sens}$ is the sensible heat.

## S3. Experimental and Computed data for Pb–Sn alloys

**Table S3.1.** Thermo-physical property data used in microscale simulations of solidifying Pb–Sn alloys by Yuan and Lee [32] and used here to calculate *Ra* an $Ra_{\mathrm{crit}}$ as a function of solid fraction $f_{\mathrm{s}}$. For their simulations, they assumed that the latent heat of fusion and heat capacity do not vary with composition.

| Property | Variable | Value |
|---|---|---|
| Initial concentration of Sn in the liquid | $C_0$ | see Table S3.2 |
| Partition coefficient | $k$ | 0.31 |
| Liquidus slope | $m_{\mathrm{L}}$ | –2.33 K wt.%$^{-1}$ |
| Composition of liquid at given solid fraction | $C_{\mathrm{L}}$ | $C_0(1-f_{\mathrm{s}})^{k-1}$ |
| Latent Heat of fusion | $H$ | 3.76x10$^{5}$ J kg$^{-1}$ |
| Specific Heat Capacity of Liquid | $C_{\mathrm{p}}^{\mathrm{liq}*}$ | 167 J kg$^{-1}$ K$^{-1}$ |
| Specific Heat Capacity of Solid | $C_{\mathrm{p}}^{\mathrm{solid}*}$ | 167 J kg$^{-1}$ K$^{-1}$ |
| Kinematic viscosity | $\nu$ | 2.47x10$^{-7}$ m$^{2}$ s$^{-1}$ |
| Solutal Expansion Coefficient | $\beta_{\mathrm{C}}$ | 5.15x10$^{-3}$ |
| Thermal Expansion Coefficient | $\beta_{\mathrm{T}}$ | 1.21x10$^{-4}$ |
| Fractional Density inversion | $\Delta\rho/\rho_0$ | $(m_{\mathrm{L}}\beta_{\mathrm{T}}+\beta_{\mathrm{C}})(C-C_0)$ |
| Change in liquid fraction with temperature† | $\dfrac{\partial f_{\mathrm{L}}}{\partial T_{\mathrm{liq}}}$ | $\dfrac{(1-f_s)^{2-k}}{m(k-1)\mathrm{C}_0}$ |
| Primary dendrite arm spacing | $\lambda_1$ | $921.7C_0^{0.134}R^{-0.261}G^{-0.354}$ |

*only one value of specific heat capacity was presented in ref. [32], which is equivalent to the common value used here.

† this provides an approximation based on the Scheil-Gulliver model for solidification. Accordingly, maximum values of *Ra* appear to differ slightly from the values given by Yuan and Lee [32].

**Table S3.2.** Cooling rate conditions for microscale simulations of directionally solidified Pb–Sn alloys by Yuan and Lee [32]. $R$ is the speed of the solidification isotherms and $G$ is the thermal gradient. The predictions of freckling behavior reported by Yuan and Lee are also given [32].

| Case | $C_0$ (wt.%) | $R$ (x10$^{-6}$ m / s$^{-1}$) | $G$ (x10$^2$ K / m$^{-1}$) | $Ra_{max}$ reported in [32] | Freckling predicted? |
|---|---|---|---|---|---|
| 1 | 10 | 330 | 3.63 | 1.2 | No |
| 2 | 10 | 41.7 | 3.63 | 25.7 | No |
| 3 | 10 | 33 | 3.63 | 36.1 | Yes |
| 4 | 10 | 3.3 | 3.63 | 1193.6 | Yes |
| 5 | 10 | 41.7 | 1.5 | 50.2 | Yes |
| 6 | 10 | 41.7 | 8 | 6.5 | No |
| 7 | 15 | 41.7 | 8 | 15.9 | No |
| 8 | 20 | 41.7 | 8 | 37.5 | Yes |
| 9 | 25 | 41.7 | 8 | 46.9 | Yes |

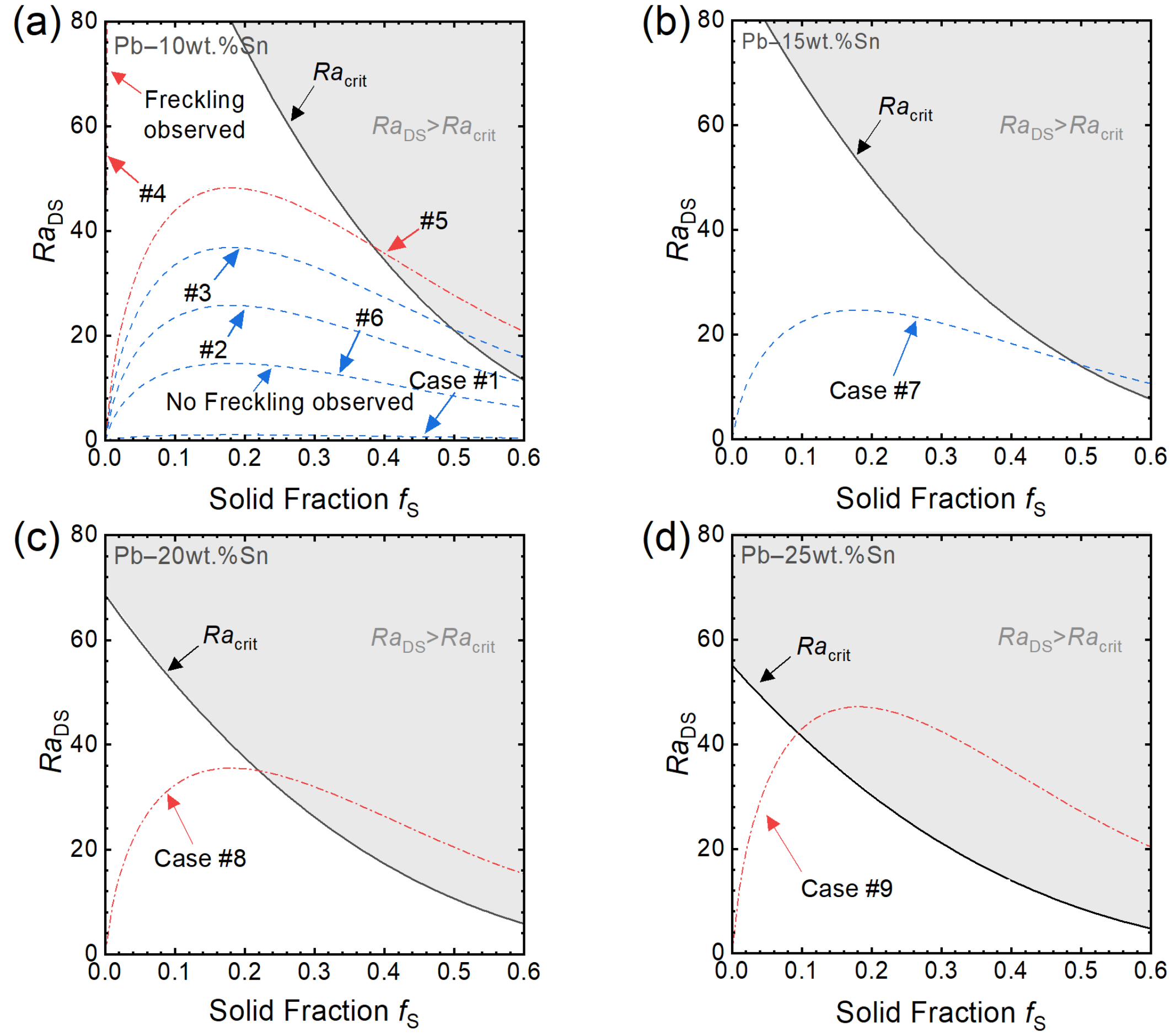

**Figure S3.1.** $Ra$ and $Ra_{crit}$ curves for (a) Pb–10wt.% Sn, (b) Pb–15wt.% Sn, (c) Pb-20wt.% Sn and (d) Pb-25wt.% Sn studied by Yuan and Lee using microscale simulations [32]. The plots are reproduced from Figure 4, but are now labelled with the specific cases (whose cooling conditions are given in Table S3.2).

## S4. Derivation of critical *Ra* number with an arbitrary inclination of the solidification front with respect to gravity

We consider a planar mushy zone advancing at a rate $\boldsymbol{R}$ due to a temperature gradient $\nabla T$ at angles θ and $\theta_2$ to $\boldsymbol{g}$, respectively (Figure S4.1). We consider only scenarios where θ and $\theta_2$ are large, as applicable to large ingot castings. In this simple treatment, we do not consider complex geometrical and transient effects, such as side-wall losses.

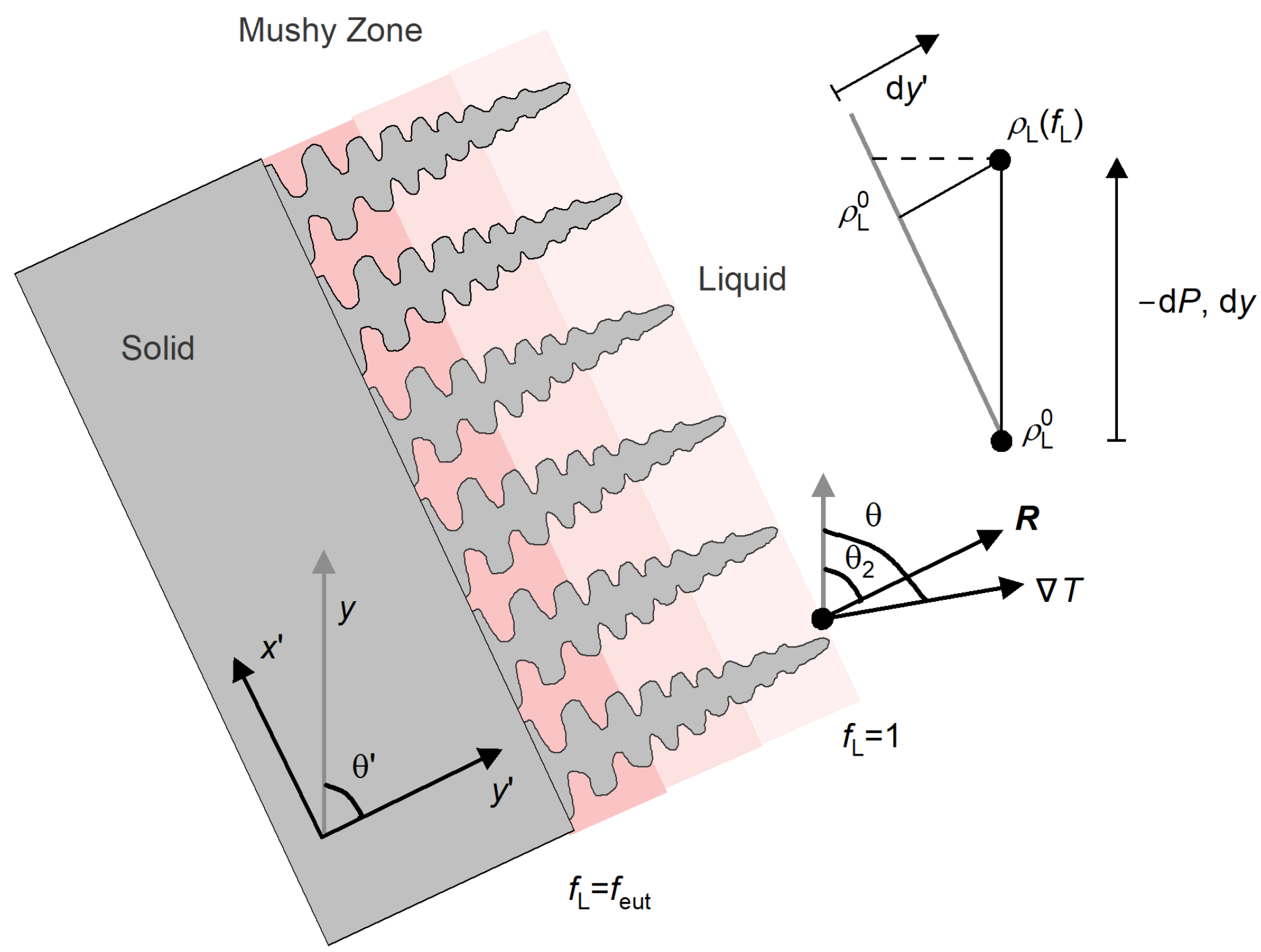

**Figure S4.1.** Schematic diagram of a planar solidification front advancing at an arbitrary orientation with respect to gravity. Dendrites are shown for illustrative purposes only.

The local-remelting condition in the mushy zone can be described by combining Eq. 11 and 12 in the main text:

$$\frac{\bar{K}}{\mu\dot{T}}(\nabla P + \rho_L \boldsymbol{g}) \cdot \nabla T > f_L \left[1 + \frac{H}{C_p(1-\beta f_L)\dot{T}} \frac{\partial f_L}{\partial t}\right]. \tag{S4.1}$$

The term $(\boldsymbol{\nabla}P + \rho_{\mathrm{L}}\boldsymbol{g}) \cdot \boldsymbol{\nabla}T$ can be evaluated as follows, assuming that the pressure gradient is due to density changes alone:

$$-(\boldsymbol{\nabla}P + \rho_{\mathrm{L}}\boldsymbol{g}) \cdot \boldsymbol{\nabla}T = (\rho(f_{\mathrm{L}}) - \rho_{\mathrm{L}}^{0})g|\boldsymbol{\nabla}T|\cos\theta \tag{S4.2}$$

The density term can be rewritten in the form:

$$-(\boldsymbol{\nabla}P + \rho_{\mathrm{L}}\boldsymbol{g}) \cdot \boldsymbol{\nabla}T = -\rho_{L,0}\left(\sum_{n}\beta_{\mathrm{T}} + \frac{\beta_{\mathrm{C},n}}{m'_{L,n}}\right)|\boldsymbol{\nabla}T|\mathrm{d}y'\cos(\theta-\theta_2)\, g|\boldsymbol{\nabla}T|\cos\theta, \tag{S4.3}$$

where $m'_{L,n}$ is the liquidus slope with respect to concentration that accounts for the type of solute. To define an effective Rayleigh number $Ra_{\mathrm{eff}}$ for tilted scenarios, we account for the fact that either a positive or negative density change across the mushy zone can cause convection at large tilt angles:

$$-(\boldsymbol{\nabla}P + \rho_{\mathrm{L}}\boldsymbol{g}) \cdot \boldsymbol{\nabla}T = -|\Delta\rho_{\mathrm{L}}|g|\boldsymbol{\nabla}T|\cos(\theta-\theta_2)\cos\theta, \tag{S4.4}$$

such that:

$$\frac{\bar{K}g}{\nu\dot{T}}\left|\frac{\Delta\rho_{\mathrm{L}}}{\rho_{\mathrm{L}}^{0}}\right||\nabla T|\cos(\theta-\theta_2)\cos\theta > f_{\mathrm{L}}\left[1 + \frac{H}{C_{\mathrm{p}}(1-\beta f_{\mathrm{L}})\dot{T}}\frac{\partial f_{\mathrm{L}}}{\partial t}\right]. \tag{S4.5}$$

In three dimensions, then the cooling rate is approximately given by:

$$\dot{T} \approx -\boldsymbol{R}\cdot\boldsymbol{\nabla}T = |\boldsymbol{R}||\boldsymbol{\nabla}T|\cos(\theta-\theta_2) \tag{S4.6}$$

such that the local remelting condition is satisfied when $Ra_{\mathrm{eff}}$ exceeds $Ra_{\mathrm{crit}}$. $Ra_{\mathrm{eff}}$ is given by:

$$Ra_{\mathrm{eff}}(\theta) = |Ra_{\mathrm{DS}}|\cos\theta = \frac{\bar{K}g\left|\frac{\Delta\rho_{\mathrm{L}}}{\rho_0}\right|}{\nu|\boldsymbol{R}|}\cos\theta \tag{S4.7}$$

and:

$$Ra_{\mathrm{crit}} = f_{\mathrm{L}}\left[1 + \frac{H}{C_{\mathrm{p}}(1-\beta f_{\mathrm{L}})\dot{T}}\frac{\partial f_{\mathrm{L}}}{\partial t}\right], \tag{S4.8}$$

The value of $Ra_{\mathrm{crit}}$ for local remelting does not explicitly depend on the orientation of the solidification front.

## S5. Experimental and Computed data for Steels

In ref. [2], two expressions were used for the secondary dendrite-arm spacing. The first was measured by Ogino *et al.* and reported in ref. [53]:

$$\lambda_2 = 123(RG)^{-0.33}\exp(-0.281C_{\mathrm{c}} - 0.175C_{\mathrm{Mn}} - 0.063C_{\mathrm{Cr}} - 0.136C_{\mathrm{Mo}} - 0.091C_{\mathrm{Ni}}), \qquad \text{(S5.1)}$$

where $C_{\mathrm{i}}$ is the composition of element *i* in wt.%. The second was reported by Won and Thomas [54]:

$$\lambda_2 = (1691 - 7209C_{\mathrm{c}})(RG)^{-0.493} \text{ for } 0 < C_{\mathrm{c}} < 0.15 \text{ wt.\%}, \qquad \text{(S5.2a)}$$

$$\lambda_2 = 143.9(RG)^{-0.36}C_C^{0.5501-1.996C_C} \text{ for } C_{\mathrm{c}} > 0.15 \text{ wt.\%}, \qquad \text{(S5.2b)}$$

Data used to calculate *Ra* and $Ra_{\mathrm{crit}}$ is provided in Table S5.1.

**Table S5.1.** Experimental data from Suzuki and Miyamoto for 27 steel compositions reported in ref. [2,13–15,45]. Their compositions are listed in ref. [2]. Values for $S_{\mathrm{crit}}$ were taken directly from ref. [13–15,43]. Values are given for *Ra*, *H*, $C_{\mathrm{p}}$, $\partial f_{\mathrm{L}}/\partial T$ and $Ra_{\mathrm{crit}}$ determined in this work using PanIron 2025 database at 30% solid fraction in the mushy zone. $\lambda_2$ is the secondary dendrite arm spacing.

| Experimental Data | | | | Calculated at 30% solid fraction | | | | |
|---|---|---|---|---|---|---|---|---|
| **Alloy composition** | **Ingot Size** | $S_{\mathrm{crit}}$ **(K s$^{-2.1}$ µm$^{1.1}$)** | **\|*R*\| (µm s$^{-1}$)** | ***Ra* at 0° tilt using $\lambda_2$ expression from Ogino *et al.* [2,53]** | ***H* (kJ kg$^{-1}$)** | $C_{\mathrm{p}}$ **(J kg$^{-1}$ K$^{-1}$)** | $\partial f_{\mathrm{L}}/\partial T_{\mathrm{liq}}$ **(K$^{-1}$)** | $Ra_{\mathrm{crit}}$ **from Eq. 15** |
| #1 | Small | 3.20 | 55±30 | 0.9 | 262 | 720 | 0.02 | 4.08 |
| #2 | Large | 1.92 | 28±10 | 1.5 | 259 | 813 | 0.013 | 3.68 |
| #3 | Large | 3.25 | 28±10 | 1.6 | 258 | 665 | 0.01 | 3.42 |
| #4 | Large | 0.35 | 28±10 | 1.2 | 245 | 802 | 0.02 | 4.95 |
| #5 | Large | 0.22 | 28±10 | 5.2 | 246 | 704 | 0.026 | 7.16 |
| #6 | Large | 0.45 | 28±10 | 2.6 | 269 | 728 | 0.026 | 7.52 |
| #7 | Large | 0.24 | 28±10 | 1.7 | 246 | 727 | 0.02 | 5.40 |
| #8 | Large | 0.09 | 28±10 | 18.8 | 243 | 676 | 0.02 | 5.71 |
| #9 | Large | 0.39 | 28±10 | 1.3 | 246 | 854 | 0.02 | 4.71 |
| #10 | Large | 0.06 | 28±10 | 5.1 | 404 | 769 | 0.031 | 12.08 |
| #11 | Large | 0.05 | 28±10 | 3.2 | 301 | 883 | 0.041 | 10.52 |
| #12 | Large | 0.14 | 28±10 | 4.0 | 243 | 897 | 0.027 | 5.74 |
| #13 | Large | 0.20 | 28±10 | 1.9 | 245 | 813 | 0.026 | 6.28 |
| #14 | Small | 0.21 | 55±30 | 1.9 | 246 | 700 | 0.026 | 7.20 |
| #15 | Small | 0.17 | 55±30 | 0.1 | 246 | 798 | 0.04 | 7.11 |
| #16 | Small | 0.10 | 55±30 | 9.1 | 245 | 714 | 0.026 | 7.05 |
| #17 | Small | 0.14 | 55±30 | 11.7 | 243 | 772 | 0.02 | 5.17 |
| #18 | Small | 0.10 | 55±30 | 7.4 | 246 | 954 | 0.04 | 7.85 |
| #19 | Small | 0.14 | 55±30 | 2.4 | 242 | 832 | 0.026 | 6.09 |
| #20 | Small | 0.07 | 55±30 | 6.2 | 476 | 970 | 0.033 | 12.13 |
| #21 | Small | 3.09 | 55±30 | 0.9 | 262 | 725 | 0.013 | 4.05 |
| #22 | Small | 0.60 | 55±30 | 4.3 | 262 | 694 | 0.012 | 2.95 |
| #23 | Small | 1.18 | 55±30 | 1.7 | 263 | 836 | 0.016 | 4.21 |
| #24 | Small | 0.85 | 55±30 | 2.5 | 263 | 792 | 0.016 | 4.41 |
| #25 | Small | 0.23 | 55±30 | 1.8 | 258 | 748 | 0.013 | 3.91 |
| #26 | Small | 2.07 | 55±30 | 2.2 | 262 | 697 | 0.013 | 4.20 |
| #27 | Small | 1.17 | 55±30 | 10.2 | 264 | 777 | 0.013 | 3.86 |

## S6. Composition dependence of the critical Suzuki number

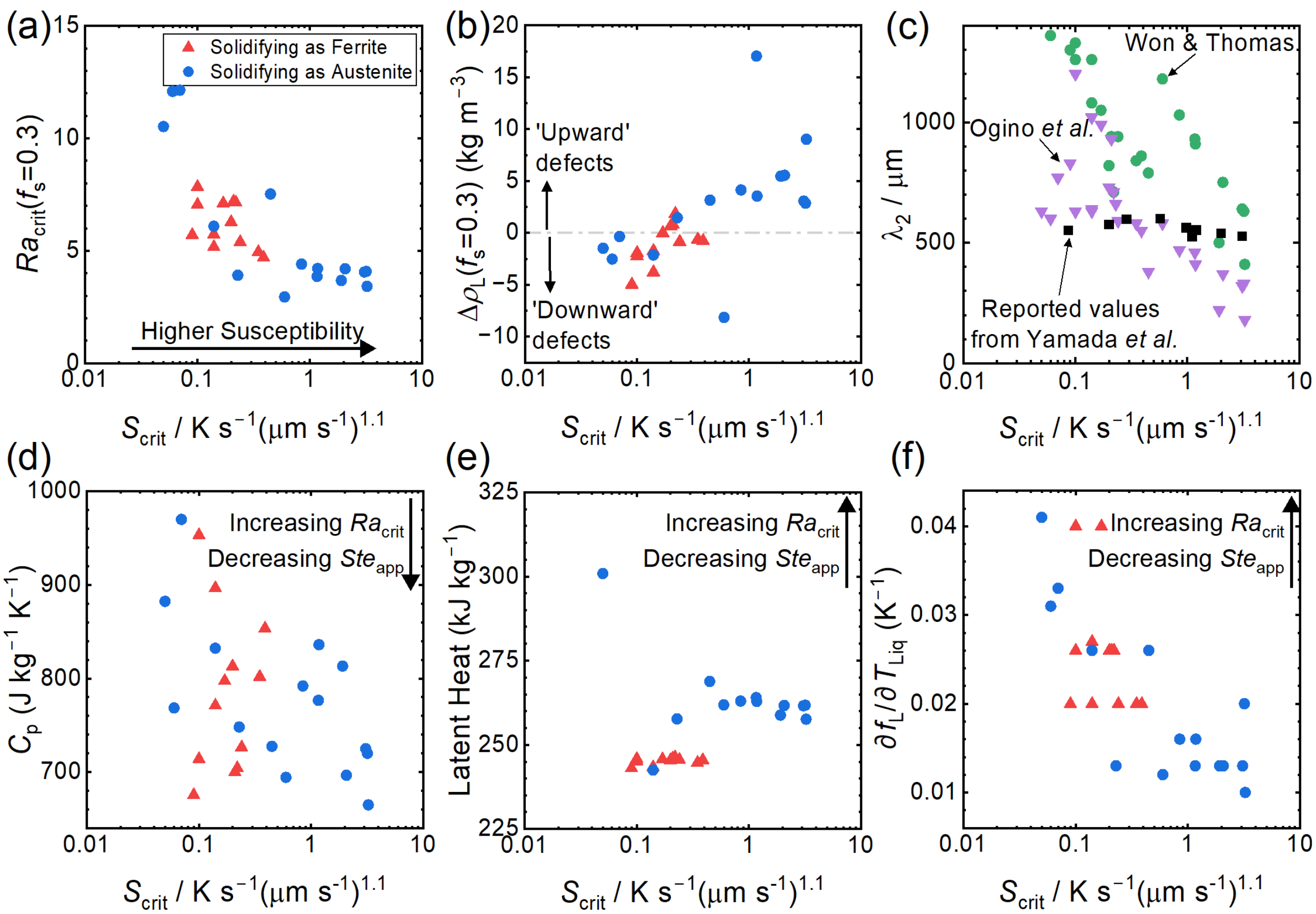

**Figure S6.1.** (a) Plot of computed $Ra_{\mathrm{crit}}$ at 30% solid fraction against experimentally measured critical Suzuki number $S_{\mathrm{crit}}$ for the 27 steel compositions considered in ref. [2]. Compositions are classified by whether they first solidify as body-centred cubic Ferrite or cubic-close packed Austenite. $Ra_{\mathrm{crit}}$ scales with the effective Stefan number $Ste_{\mathrm{eff}}$. (b) Plot of computed density inversion $\Delta\rho_{\mathrm{L}}$ at 30% solid fraction against $S_{\mathrm{crit}}$. (c) Plot of secondary dendrite arm spacing predicted from the two models adopted in ref. [2] against $S_{\mathrm{crit}}$. Experimental measurements from Yamada *et al.* for selected experiments are also shown [45]. Reproduced from the main text for easy comparison with other trends. (c-f) Plots of computed values of specific heat capacity, latent heat and $\partial f_{\mathrm{L}}/\partial T_{\mathrm{liq}}$ against $S_{\mathrm{crit}}$.

**Additional references**